\documentclass[aps,prd,superscriptaddress,nofootinbib,amsfonts,amssymb,amsmath,notitlepage,twocolumn]{revtex4-1}
\usepackage[utf8x]{inputenc}
\usepackage{graphicx}
\usepackage{float}
\usepackage{wrapfig}
\usepackage{bm}
\usepackage{color}
\usepackage{comment}
\usepackage{xcolor}
\usepackage[normalem]{ulem}
\usepackage{hyperref}
\hypersetup{
colorlinks=true,
citecolor=blue,
citebordercolor=red,
linktoc=all,
linkcolor=blue,
urlcolor=blue
}
\usepackage{amsmath,amsfonts,amsthm,amssymb,mathrsfs,latexsym,paralist}

\catcode`\@=11
\def\@versim#1#2{\vcenter{\offinterlineskip
\ialign{$\m@th#1\hfil##\hfil$\crcr#2\crcr\sim\crcr } }}
\catcode`\@=12

\newcommand{\Slash}[1]{{\ooalign{\hfil/\hfil\crcr$#1$}}}

\newcommand{\p}{\partial}

\newcommand{\bp}{\begin{pmatrix}}
\newcommand{\ep}{\end{pmatrix}}
\newcommand{\nn}{\nonumber\\}

\newcommand{\df}{\text{d}}

\newcommand{\bs}[1]{\boldsymbol}

\newcommand{\Tr}{{\rm Tr}\,}
\newcommand{\tr}{{\rm tr}\,}

\newcommand{\pmat}[1]{\begin{pmatrix}#1\end{pmatrix}}

\newcommand{\commentmute}[1]{}

\graphicspath{{./Figs/}}

\begin{document}
\newbox{\ORCIDicon}
\sbox{\ORCIDicon}{\large
                  \includegraphics[width=0.8em]{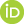}}

\title{On the phase structure of extra-dimensional gauge theories with fermions
}

%%%%%%%%%%%%%%%%%%%%%%%%%%%%%%%%%%%%
%%%%%%%%%%%%%%%%%%%%%%%%%%%%%%%%%%%%
\author{\'Alvaro \surname{Pastor-Guti\'errez}\,\href{https://orcid.org/0000-0001-5152-3678}{\usebox{\ORCIDicon}}}
%\email{pastor@mpi-hd.mpg.de}
\affiliation{Max-Planck-Institut f\"ur Kernphysik P.O. Box 103980, D 69029, Heidelberg, Germany}
%\affiliation{Institut f\"ur Theoretische Physik, Universit\"at Heidelberg, Philosophenweg 16, 69120 Heidelberg, Germany}
%%%%%%%%%%%%%%%%%%%%%%%%%%%%%%%%%%%%
\author{Masatoshi \surname{Yamada}\,\href{https://orcid.org/0000-0002-1013-8631}{\usebox{\ORCIDicon}}}
%\email{yamada@jlu.edu.cn}
\affiliation{Center for Theoretical Physics and College of Physics, Jilin University, Changchun 130012, China}
%%%%%%%%%%%%%%%%%%%%%%%%%%%%%%%%%%%%
%%%%%%%%%%%%%%%%%%%%%%%%%%%%%%%%%%%%

\begin{abstract}
We study the phase structure of five-dimensional Yang-Mills theories coupled to Dirac fermions. In order to tackle their non-perturbative character, we derive the flow equations for the gauge coupling and the effective potential for the Aharonov-Bohm phases employing the Functional Renormalisation Group. We analyse the infrared and ultraviolet fixed-point solutions in the flow of the gauge coupling as a function of the compactification radius of the fifth dimension. We discuss various types of trajectories which smoothly connect both dimensional limits. Last, we investigate the phase diagram and vacuum structure of the gauge potential for different fermion content. 
\end{abstract}
\maketitle

\section{Introduction}

Many quantum field theories (QFTs), like the Standard Model of particle physics (SM), are formulated as non-Abelian gauge theories. Understanding the phase structure and dynamics of this family of theories is one of the central tasks carried on by theoretical and phenomenological efforts. Whereas our universe is well-described by a four-dimensional spacetime system,  the possible emergence of extra spacetime dimensions at higher energy scales is not yet excluded. These could be used to solve long-standing open questions in fundamental physics such as the nature of the electroweak scale~\cite{Gildener:1976ai,Weinberg:1978ym}. 
For example, extra-dimensional gauge theories have been studied as models for Gauge-Higgs unification~\cite{Fairlie:1979zy,Manton:1979kb,Forgacs:1979zs,Hosotani:1983xw,Hosotani:1983vn,Hosotani:1988bm,Davies:1987ei,Davies:1988wt,Antoniadis:1993jp,Antoniadis:2001cv} where one of the components in the gauge field is identified with the Higgs field. This idea has been extended to more realistic setups beyond the SM (BSM). See e.g. Refs.~\cite{Pomarol:1998sd,Kubo:2001zc,Agashe:2004rs,Medina:2007hz,Hosotani:2008tx,Funatsu:2019fry,Funatsu:2020znj,Haba:2002vc,Lim:2007jv,Kojima:2011ad,Hosotani:2015hoa,Yamatsu:2015oit,Furui:2016owe,Angelescu:2021nbp,Angelescu:2022obm,Kojima:2023mew}.  

Four-dimensional ($D=4$) non-Abelian gauge theories such as Quantum Chromodynamics (QCD) have been thoroughly studied with various QFT methods. One of their noteworthy features is asymptotic freedom \cite{Gross:1973id,Politzer:1973fx} in the ultraviolet (UV) limit which allows us to define them as continuum QFTs. In the infrared (IR) limit, the theories evolve into a strongly interacting sector with a rich phase structure and non-trivial phenomena such as dynamical chiral symmetry breaking and colour confinement. In the many-flavour limit, additional non-trivial IR fixed-point solutions can appear. In $D=4$, these are known as known as Caswell-Banks-Zaks (CBZ) fixed points~\cite{Caswell:1974gg,Banks:1981nn}.

Conversely, extra-dimensional gauge theories have not been as studied due to their perturbatively non-renormalisable character. Nevertheless, they show several attractive features for example evidence for the existence of a non-trivial UV fixed-point solution providing the theory with UV completeness~\cite{Peskin:1980ay,Morris:2004mg,Irges:2018gra,Irges:2020nap,DeCesare:2021pfb,Gies:2003ic,Pastor-Gutierrez:2022rac} and hence, asymptotic safety. Moreover, in this scenario, the theory although perturbatively not renormalisable is non-perturbatively renormalisable. 

An important object to properly understanding the phase structure of gauge theories is the gauge potential. A key phenomenon in gauge theories, especially on compactified spacetimes e.g. $\mathbb R^4\times S^1$, is the Hosotani mechanism~\cite{Hosotani:1983xw,Hosotani:1983vn,Hosotani:1988bm} which is analogous to the Aharonov-Bohm (AB) effect in quantum mechanics. Here, the extradimensional component of the five-dimensional gauge field shows a non-trivial background which, depending on its configuration, can lead to the breaking of the gauge symmetry into one of its subgroups. This dynamical process can be used to formulate the SM electroweak and Higgs sectors in a more minimal manner where all degrees of freedom are embedded into a five-dimensional gauge theory.

In this work, we study the phase structure of five-dimensional Yang-Mills theories coupled with Dirac fermions transforming in different representations of the gauge group. To continuously connect with the well-known four-dimensional limit we formulate the theory in a compactified manner on $\mathbb R^4\times S^1$ spacetime. To tackle the non-perturbative character of these theories, we employ the functional renormalisation 
group (fRG)~\cite{Wetterich:1992yh,Morris:1993qb,Ellwanger:1993mw,Morris:1998da,Berges:2000ew,Aoki:2000wm,Bagnuls:2000ae,Polonyi:2001se,Pawlowski:2005xe,Gies:2006wv,Delamotte:2007pf,Sonoda:2007av, Rosten:2010vm,Kopietz:2010zz,Braun:2011pp,Dupuis:2020fhh} which has proven its applicability in similar frameworks. We employ the background field approximation~\cite{Reuter:1993kw,Reuter:1992uk,Reuter:1993nn,Reuter:1994sg,Wetterich:2017aoy} and derive the flow equations employing heat-kernel techniques. For the fixed-point analysis, we help ourselves by employing perturbative results for the anomalous dimensions which include all fermionic contributions up to three loops. While the truncation employed in this work grants access to the qualitative features, it fails to quantitatively address the strong dynamics and the associated phenomena such as colour confinement and chiral symmetry breaking. Therefore, we study the behaviour of the gauge coupling and configurations of the background gauge fields depending on fermion numbers.

This work is organised as follows. After introducing the theoretical framework and the effective action formalism in Section~\ref{sec: Model and setups}, in Section~\ref{sec: Gauge coupling and phase structure} we derive the flow equation for the gauge coupling and study the fixed-point solutions as a function of the compactification radius and fermion content.  In Section~\ref{sec: Potential of AB phases}, we derive the flow of the effective potential and study the AB phase diagram as a function of the compactification radius and fermion masses for different number of fundamental and adjoint fermions. Section~\ref{sec: Summary} is devoted to summarizing the results in this work and their relevance.

\section{Theoretical framework}
\label{sec: Model and setups}

In this work, we employ the functional Renormalisation Group (fRG) to tackle the non-perturbative character of five-dimensional gauge theories. In QFT, the fRG is formulated as a functional partial differential equation describing the change of the effective average action (one-particle irreducible generating functional) $\Gamma_k$ for the varying energy scale $k$.
The scale dependence of $\Gamma_k$ obeys the Wetterich equation~\cite{Wetterich:1992yh} whose form is given by
\begin{align}
	\p_t \Gamma_k =\frac{1}{2}\Tr \left[ \left( \Gamma_k^{(2)} + \mathcal R_k \right)^{-1}\p_t \mathcal R_k \right]\,,
	\label{eq: flow equation}
	\end{align}
where $\p_t=k\,\p_k$ is the dimensionless scale and $\Tr$ is the functional trace acting on all spaces in which the fields are defined.
Here, $\Gamma_k^{(2)}$ is the full two-point correlation function obtained by performing the second-order functional derivative for $\Gamma_k$ and $\mathcal R_k$ is the regulator function realizing the coarse graining process within the path integral.

We are especially interested in the impact of fermionic fluctuations on the gauge coupling and the non-trivial background field of the gauge field in five dimensional (Euclidean) spacetime. For this purpose, we make the following ansatz for the effective action $\Gamma_k$,
\begin{align}
\Gamma_k &=
 \Gamma_\text{gauge} + \Gamma_\text{fermion} + S_\text{gf} + S_\text{gh}+\Gamma_\text{AB} \,.
\label{eq: effective action}
\end{align}
Here, $\Gamma_\text{gauge}$ is the effective action for the $SU(N_c)$ Yang-Mills gauge field
\begin{align}
\Gamma_\text{gauge}&=\frac{Z_k}{4g^2}\,\int \df^5x\,F^a_{MN}F^{aMN}\,,
\label{eq: effective action for gauge field}
\end{align}
which  in the current approximation has been truncated at lowest order of gauge invariant operators. Here, $F_{MN}^a=\p_M A_N^a -\p_N A_M^a -f^{abc}A^b_M A^c_N$ is the field strength of $A_M^a$ with $f^{abc}$ the structure constants of $SU(N_c)$ and $Z_k$ is the field renormalisation factor for the gauge field. Note that the squared gauge coupling has a canonical mass dimension of
\begin{align}
 [g^2]=-1\,,
\end{align}
making apparent the power counting non-renormalisable character of the theory. Here and hereafter, capital Latin characters $M,\,N,\cdots(=0,\ldots, 3,5)$ stand for Lorentz indices in the five dimensional spacetime, while four dimensional Lorentz indices are denoted by Greek letters $\mu, \nu,\cdots(=0,\ldots, 3)$ and indices for the fundamental and adjoint representations of $SU(N_c)$ are denoted by small Latin characters $i,j,\cdots(=N_c)$ and $a,b,\cdots(=N_c^2-1)$, respectively. 

We consider fermion fields in both fundamental and adjoint representations of $SU(N_c)$, which are denoted by $\psi=(\psi)_i$ and $\chi=(\chi_a\tau^a)_{ij}$, respectively, where $\tau^a$ are generators of $SU(N_c)$ in the fundamental representation.
Note that the normalisation for these generators is chosen to be $\tr(\tau^a\tau^b)=\frac{1}{2}\delta^{ab}$.  The truncated  effective action reads
\begin{align}
\Gamma_\text{fermion}&=\int \df^5x \,\bar\psi  \left(iZ_\psi\Gamma^M \nabla_M + m_\text{f} \right) \psi \notag\\
 &\hspace{.4cm}+ \int \df^5x \,\bar\chi_a \left(iZ_\chi\Gamma^M D_M + m_\text{ad} \right)^{ab} \chi_b\,,
 \label{eq: effective action for spinors}
\end{align}
where $\Gamma_M$ are the Dirac matrices in five dimensional spacetime and are defined by  those in four dimensions,
\begin{align}
&\Gamma_M = \gamma_\mu && (\text{for $M=\mu=0,\ldots,3$})\,,\\[1ex]
&\Gamma_5 = \gamma_5 && (\text{for $M=5$})\,. \label{eq:gamma5inD5}
\end{align}
satisfying the Clifford algebra $\{ \Gamma_M, \Gamma_N\} =2 \delta_{MN}$. The covariant derivatives acting on spinor fields are given respectively by 
\begin{align}
\nabla_M \psi&= \p_M \psi -iA_M\psi\,,
\label{eq: CD for fund spinor}
\\[1ex]
D_M\chi&= \p_M\chi -i[A_M, \chi] \,,
\label{eq: CD for ad spinor}
\end{align}
where the fields in adjoint representation are $\chi=\chi_a\tau^a$ and $A_M=A_M^a\tau^a$. 

Additionally in the gauge sector, the gauge fixing and ghost actions associated to the gauge symmetry read
\begin{align}
S_\text{gf}&= \frac{Z_k}{\xi}\int \df^5x\, \tr[\mathcal F]^2\,,
\label{eq: gauge fixing action}
\\[1ex]
S_\text{gh}&= Z_\text{gh}\int \df^5 x\, \tr\left[ \bar c \,{\mathcal M} c\right]\,,
\label{eq: ghost action}
\end{align}
with $\xi$ being the gauge fixing parameter and $Z_\text{gh}$ the ghost field renormalisation factor. Here, $c=\tau^a c^a$ and $\bar c=\tau^a\bar c^a$ denote the ghost and anti-ghost fields, respectively and tr acts on $SU(N_c)$ spaces. 
In this work, we intend to investigate the non-trivial background generation of $A_M^a$, so that it is convenient to introduce the gauge fixing function $\mathcal F$ such that some interactions involving both the background and fluctuation fields in Eq.~\eqref{eq: effective action for gauge field} are appropriately cancelled. To this end, assuming the linear splitting $A_M^a=\bar A_M^a + a_M^a$, we give a type of the $R_\xi$-gauge fixing function for $\mathcal F$, namely
\begin{align}
\mathcal F &= \delta^{\mu\nu}\bar D_\mu a_\nu + \xi \bar D_5 a_5\,,
\label{eq: gauge fixing function}
\end{align}
where bar on the field stands for the background gauge field and then $\bar D_\mu a_\nu = \p_\mu a_\nu - i[\bar A_\mu, a_\nu]$ is the covariant derivative with the background gauge field $\bar A_M^a$ and the fluctuation $a_M$. Once $\mathcal F$ is fixed, the derivative operator $\mathcal M$ acting on the ghost fields is automatically determined by the BRST quantization procedure~\cite{Kugo:1979gm} to be
\begin{align}
{\mathcal M} &=\delta^{\mu\nu} \bar D_\mu D_\nu + \xi \bar D_5 D_5\,.
\end{align}
In this study, we choose the Feynman gauge $\xi=1$ to appropriately cancel the $\bar A_5$ contributions appearing in the gauge and ghosts modes. Hereafter, we provide all formulas in this gauge.

When a five dimensional spacial direction is compactified such that $\mathbb R^4\times S^1$ in which the circle radius is denoted by $R\geq |x_5|$, its Fourier modes are discretized and known as the Kaluza-Klein (KK) modes. More specifically, the fluctuation fields for gauge and ghost fields ($\varphi \equiv a_M, c,\bar c$) in position space are given by
\begin{align}
\varphi(x,x_5)&= \sum_{n=-\infty}^\infty \varphi^{(n)}(x) \frac{e^{inx_5 /R}}{\sqrt{2\pi R}}\,,
\label{eq: KK expansion for varphi}
\end{align}
while the fermionic fields, $\psi$, $\chi$, are expanded as
\begin{align}
\psi(x,x_5) = \sum_{n=-\infty}^\infty \psi^{(n)}(x) \frac{e^{i(n+\beta_{\rm f}/2\pi)(x_5 /R)}}{\sqrt{2\pi R}}\,,\\
\chi(x,x_5) = \sum_{n=-\infty}^\infty \chi^{(n)}(x) \frac{e^{i(n+\beta_{\rm ad}/2\pi)(x_5 /R)}}{\sqrt{2\pi R}}\,.
\end{align}
These KK expansions are associated to the boundary conditions
\begin{align}
&\varphi(x,x_5+2\pi R)= \varphi(x,x_5) \,, \\
&\psi(x,x_5+2\pi R) = e^{i\beta_{\rm f}}\psi(x,x_5)\,,\\ 
&\chi(x,x_5+2\pi R) = e^{i\beta_{\rm ad}}\chi(x,x_5)\,,
\end{align}
where $\beta_\text{f}\in [0,2\pi)$ and  $\beta_\text{ad}\in [0,2\pi)$.

For the compactification $\mathbb R^4\times S^1$, we consider the form 
\begin{align}
&\bar A_\mu^a =0\,,&
&\bar A_5^a = \frac{\vartheta^a}{2\pi R}\,.
\label{eq: background field of A}
\end{align}
for the background gauge field, which is written in a unified manner as $\bar A_M^a=\vartheta^a \delta_M^5/(2\pi R)$.
The dimensionless parameters $\vartheta^a$ are the so-called AB phases. The use of the $SU(N_c)$ generators in the fundamental representation gives the other parametrization in a diagonal form
\begin{align}
(\bar A_5)_{ij} 
= \frac{1}{2\pi R}\pmat{
\theta_1 & & \\
 & \ddots & \\
 & & \theta_{N_c}
}
= \frac{\theta_i}{2\pi R} \delta_{ij}\,,
\end{align}
whose AB phases $\theta_i$ satisfy $\sum_{i=1}^{N_c}\theta_{i} =0$ (mod $2\pi$) due to the traceless property for the generators of $SU(N_c)$.
This parametrization can be obtained by an $SU(N_c)$ transformation $\mathcal U$ such that $\mathcal U^\dagger \vartheta^a\tau^a\mathcal U = \text{diag}(\theta_1,\cdots, \theta_{N_c})$.
Inserting the KK expansions exhibited in Eqs.~\eqref{eq: KK expansion for varphi}, the covariant derivative $\bar D_5^2$ acting on $\varphi$ and the background field \eqref{eq: background field of A} gives rise to the effective``masses" of $\varphi_{ij}^{(n)}=(\varphi_{a}^{(n)}\tau^a)_{ij}$:
\begin{align}
M_{ij,n}^2 =\frac{1}{R^2}\left( n -\frac{\theta_i-\theta_j}{2\pi}\right)^2\,.
\end{align}
We note here that the difference $\theta_i-\theta_j$ originates from the commutator involved in the covariant derivative acting on the fields: $(\bar D_M \varphi)_{ij}=(\bar D_M \varphi^a \tau^a)_{ij}=\p_M \varphi_{ij} - i[\bar A_5, \varphi]_{ij}= \p_M \varphi_{ij} -\frac{i}{2\pi R} (\theta_i -\theta_j)\varphi_{ij}$. It is clear now that even for the zero KK mode $n=0$, the gauge field acquires a finite mass from the AB phases for finite values of $R$. Furthermore, the phenomenon that the gauge symmetry is spontaneously broken by finite AB phase is known as the Hosotani mechanism~\cite{Hosotani:1983xw,Hosotani:1983vn,Hosotani:1988bm}. See also Refs.~\cite{Davies:1987ei,Davies:1988wt}. In the five dimensional spacetime limit $R\to \infty$, all KK modes become massless (for vanishing AB phases) and thus are dynamical degrees of freedom. In the four dimensional limit ($R\to0$), all KK modes are infinitely massive and only the massless zero KK mode ($n=0$) $\varphi^{(0)}$ contributes to the quantum dynamics in the system. 
The covariant derivatives acting on the fermionic fields given in Eqs.~\eqref{eq: CD for fund spinor} and \eqref{eq: CD for ad spinor} now read
\begin{align}
\nabla_M \psi&= \p_M \psi -ia_M \psi  - \bar A_5 \psi\,,\\[1ex]
D_M\chi &= \p_M\chi  -i [a_M,\chi]  -i[\bar A_5, \chi]  \,.
\end{align}
Thus, together with the Dirac masses in Eq.~\eqref{eq: effective action for spinors}, the effective squared masses for the fermionic fields $\psi_i^{(n)}$ and $\chi_{ij}^{(n)}=(\chi^{(n)}_a\tau^a)_{ij}$ are found to be
\begin{align}
M_{\text{f},i,n}^2 &= m_{\rm f}^2 + \frac{1}{R^2}\left( n +\frac{\beta_{\rm f}-\theta_i}{2\pi} \right)^2\,,\\[1ex]
M_{\text{ad},ij,n}^2 &= m_{\rm ad}^2 + \frac{1}{R^2}\left( n +\frac{\beta_{\rm ad}- (\theta_i-\theta_j)}{2\pi} \right)^2\,.
\end{align}

Finally, $\Gamma_\text{AB}$ contains the effective action for AB phases,
\begin{align}
\Gamma_\text{AB} = \int \df^5x \, V_\text{5D}(\theta_H)
=\int \df^4x \, (2\pi R)V(\theta_H)\,,
\end{align}
where $\theta_H=\{\theta_1,\cdots, \theta_{N_c}\}$ is a set of AB phases and we have performed the integral for the fifth direction of spacetime $\int^{2\pi R}_0 \df x_5=2\pi R$.

For the total effective action \eqref{eq: effective action}, the Wetterich equation \eqref{eq: flow equation} is expressed diagrammatically as
\begin{align}
\p_t \Gamma_k &= \frac{1}{2}\vcenter{\hbox{\includegraphics[width=13mm]{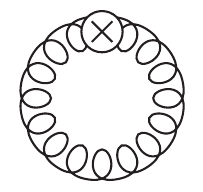}}} 
+\frac{1}{2}\vcenter{\hbox{\includegraphics[width=13mm]{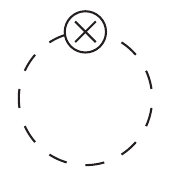}}} 
- \vcenter{\hbox{\includegraphics[width=13mm]{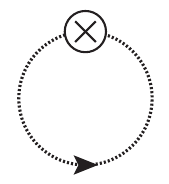}}}\nn
&\quad 
- N_{\rm f}\vcenter{\hbox{\includegraphics[width=13mm]{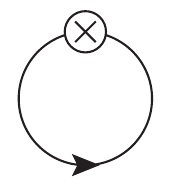}}} 
- N_{\rm ad} \vcenter{\hbox{\includegraphics[width=13mm]{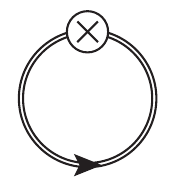}}} \,,
\label{eq:diagramatic flow}
\end{align}
where curly, dashed and dotted lines represent the propagators of the 4 dimensional gauge ($M=0,\ldots,3$), the extra dimensional gauge field component ($M=5$) and ghost fields, respectively. Single and double solid lines stand for those of fundamental and adjoint fermion fields, respectively, and the crossed circles denote the insertion of $\p_t \mathcal R_k$.

A crucial object for deriving the beta functions from the Wetterich equation is the regulated propagators (inverse two-point functions). 
For the gauge and ghost fields these have been computed in Ref.~\cite{Pastor-Gutierrez:2022rac}. The second-order functional derivatives for the fermion fields in the effective action \eqref{eq: effective action for spinors} on the background field \eqref{eq: background field of A} with respect to the pairs of ($\psi_i^{(n)}$, $\bar\psi_i^{(n)}$), and ($\chi_{ij}^{(n)}$, $\bar\chi_{ij}^{(n)}$) yield
\begin{align}
&\Gamma^{(2)}_{\bar \psi \psi}\Big|_{a_M=0} =  i\Slash p + \frac{i}{R}\left(n + \frac{\beta_\text{f} - \theta_i }{2\pi} \right)\gamma^5 + m_\text{f}\,,\\[1ex]
&\Gamma^{(2)}_{\bar \chi \chi}\Big|_{a_M=0} = i\Slash p + \frac{i}{R}\left(n + \frac{\beta_\text{ad}- (\theta_i-\theta_j)}{2\pi} \right)\gamma^5  + m_\text{ad}\,,
\end{align}
where $\Slash p= \gamma^\mu p_\mu$ with $p_\mu$ being the four dimensional momenta. 

The cutoff regulator function is introduced as $\mathcal R_k(p)=i\Slash p\, r_k(p^2/k^2)$ which replaces $i\Slash p$ in the inverse two-point functions to $i\Slash p(1+ r^F_k(p^2/k^2))$. In this work, we use the Litim-type cutoff function~\cite{Litim:2001up}, i.e., $r^F_k(x) = \left( 1/\sqrt{x} - 1\right) \theta(1-x)$ where $\theta(1-x)$ is the step function. The denominators of the regulated propagators for $\psi_i^{(n)}$ and $\chi_{ij}^{(n)}$ take the forms of $P^F_k(p)+M_{\text{f},i,n}^2$ and $P^F_k(p)+M_{\text{ad},i,n}^2$, respectively, in which we defined $P^F_k(p) = p^2 (1 + r^F_k(p^2/k^2))^2$. Note that for the bosonic fields $\varphi_{ij}^{(n)}$, we employ the cutoff function $\mathcal R_k(p)$ such that the four dimensional squared momentum $p^2$ is replaced to $P^B_k(p)=p^2(1+ r_k^B(p^2/k^2))$ with $r_k^B(x)=(1/x -1)\theta(1-x)$. For the momentum integral which is a part of the functional trace in the Wetterich equation, both $P_k^F$ and $P_k^B$ turn to $k^2$ because of $\theta(1-p^2/k^2)$. 

In the following sections, we derive the flow equations for $V(\theta_H)$ and the gauge coupling $g^2$. To this end, we use the heat-kernel method as previously developed in \cite{Pastor-Gutierrez:2022rac}. For obtaining the potential of AB phases, we project out $\bar A_5^a$ (or equivalently $\vartheta^a$) from loop effects. On the other hand, the beta function for the gauge coupling is extracted by projecting on $\bar F_{a\mu\nu} \bar F^{a\mu\nu}$ which is the field strength for $\bar A_\mu^a$. Hence, to derive the flow equation for the gauge coupling, we assume the background field of the gauge field such that $\bar F^a_{MN} \bar F^{aMN}=\bar F^a_{\mu\nu} \bar F^{a\mu\nu}$ for which the effective action \eqref{eq: effective action for gauge field} reads
\begin{align}
\frac{Z_k}{4g^2}\int \df^5x \,\bar F^a_{MN}\bar F^{a MN} = \frac{2\pi R Z_k}{4g^2} \int \df^4x\, \bar F^a_{\mu\nu} \bar F^{a\mu\nu}\,.
\end{align}
The scale derivative on the left-hand side of the Wetterich equation acts on $Z_k$, while we extract the operator $\int \df^4x \,\bar F^a_{\mu\nu} \bar F^{a\mu\nu}$ with loop coefficients and the threshold functions from the right-hand side. Note that $g^2_\text{4D}=g^2/(2\pi R)$ is identified with the dimensionless gauge coupling in four dimensional spacetime.

\section{Flow of the gauge coupling and fixed points} \label{sec: Gauge coupling and phase structure}

In this section, we derive the flow of the gauge coupling in the background field approximation and analyse the fixed-point solutions as a function of the gauge group, fermion content and the effective dimension. Additionally, we discuss interdimensional trajectories which smoothly connect all UV and IR solutions found.

\subsection{Flow equation for the gauge coupling}

We first deal with the gauge coupling flow whose detailed derivation except for fermionic loop effects has been discussed in Ref.~\cite{Pastor-Gutierrez:2022rac}. Here, we exhibit the final form of the flow equation. The running of the gauge coupling arises from that of $Z_k$ for which we define the anomalous dimension of the gauge field as
\begin{align}
\eta_g&=-\frac{\p_t Z_k}{Z_k} = -\frac{1}{(4\pi)^2}\frac{Z_k^{-1} g^2}{2\pi  R} \Bigg\{ \nn[1ex]
&\hspace{1.75cm}+7N_c \left(1-\frac{19\eta_g}{42} + \frac{\eta_{\rm gh}}{21}\right) J(\bar R; 0) \nn[1ex]
&\hspace{1.75cm}-\frac{4}{3}N_{\rm f} (1- \eta_{\rm f}) J(\bar R; \bar m_{\rm f})\nn[1ex]
&\hspace{1.75cm}- \frac{8}{3} N_c N_{\rm ad}(1 -\eta_{\rm ad}) J(\bar R; \bar m_{\rm ad}) \Bigg\}\,,
	\label{eq: anomalous dimension of etag}
\end{align}
where the anomalous dimensions for each of the remaining fields read $\eta_\text{gh}=-\p_t Z_\text{gh}/Z_\text{gh}$, $\eta_\text{f}=-\p_t Z_\text{f}/Z_\text{f}$ and $\eta_\text{ad}=-\p_t Z_\text{ad}/Z_\text{ad}$. The threshold functions for massive fields read
\begin{align}
	J(\bar R; \bar m)
	&=\sum_{n=-\infty}^\infty \frac{1}{1 + \left(\frac{n}{\bar R}\right)^2 + \bar m^2}\nn[1ex]
	&= \frac{\pi \bar R\,\coth (\sqrt{1+\bar m^2}\,\pi \bar R)}{\sqrt{1+\bar m^2}}\,,
	\label{eq: threshold function J}
\end{align}
with dimensionless quantities $\bar R=R\,k$ and $\bar m^2=m^2/k^2$.
The first term in the curly brackets on the right-hand side of Eq.~\eqref{eq: anomalous dimension of etag} corresponds to the contributions from gauge and ghost fields, while the second and third, to quantum corrections from $N_\text{f}$ fundamental fermions and $N_\text{ad}$ adjoint ones, respectively.

%%%%%%%%%%%%%%%
%%%%%%%%%%%%%%%
\begin{figure}
	\centering
	\includegraphics[width=.47\textwidth]{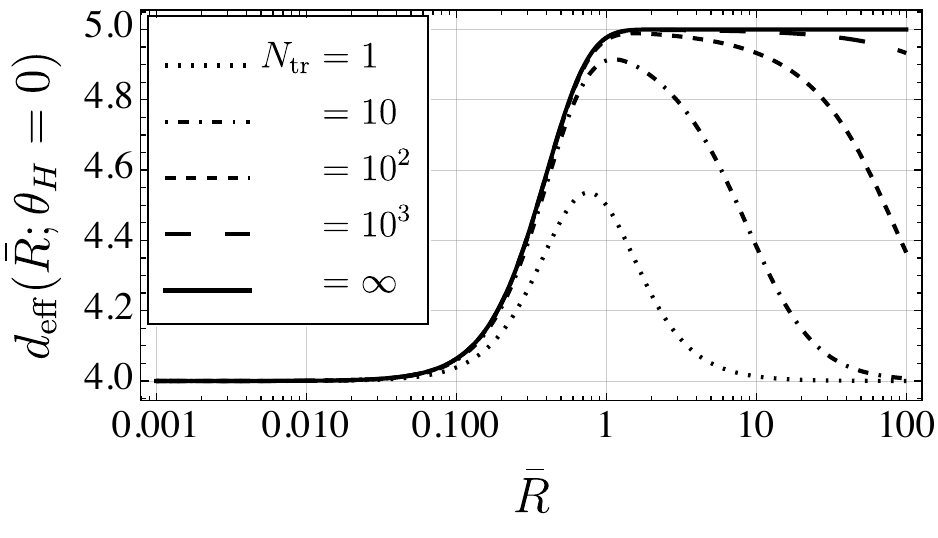} \vspace{-.5cm}
	\caption{Effective dimension for a trivial gauge background and as a function of the dimensionless compactification radius for different truncations of the KK-mode tower ($N_{\rm tr}$). Figure adapted from Ref.~\cite{Pastor-Gutierrez:2022rac}. }
	\label{Fig:effective dimension}
\end{figure}
%%%%%%%%%%%%%%%
%%%%%%%%%%%%%%%

Let us now define the dimensionless renormalised gauge coupling~\cite{Pastor-Gutierrez:2022rac} as
\begin{align}
	\tilde g^2=  \frac{Z^{-1}_kg^2}{2\pi  R} J(\bar R;0) =Z^{-1}_k g^2_\text{4D} J(\bar R;0)\,.
	\label{eq: dimensionless gauge coupling}
\end{align}
Acting $\p_t$ on this yields the flow equation for $\tilde g$ as 
\begin{align}
	\p_t \tilde g^2=\beta_{\tilde g^2}  = \left( d_\text{eff}(\bar R)-4 + \eta_g  \right) \tilde g^2 \,.
	\label{eq: beta function gtilde}
\end{align}
Here, we have defined the effective dimension
\begin{align}
	d_\text{eff}(\bar R) = 4+ \frac{\df \log J(\bar R;0)}{\df \log\bar R}\,,
	\label{eq: effective dimension}
\end{align}
as a function of $\bar R$ \cite{Kubo:1999ua,Pastor-Gutierrez:2022rac}. As depicted in Fig.~\ref{Fig:effective dimension}, the limit $\bar R\to \infty$ corresponds to $D=5$ and gives $d_\text{eff}(\infty)=5$, while in the $D=4$ limit ($\bar R\to 0$), we have  $d_\text{eff}(0)=4$. Hence, Eq.~\eqref{eq: effective dimension} allows us to smoothly and continuously interpolate the spacetime dimensionality of the system depending on the compactification radius. 

The anomalous dimension $\eta_g$ appears on both sides of Eq.~\eqref{eq: anomalous dimension of etag} and hence can easily be solved so as to be
\begin{align}\label{eq: anomalous dimension resumed}
	\eta_g &= \frac{ -\frac{ {\tilde g}^2}{(4 \pi)^2} \left( \frac{N_c}{3} \left(22-1\right)- \frac{4}{3} N_{\rm f} - \frac{8}{3}N_{\rm ad}N_c \right)}{1 -\frac{{\tilde g}^2}{(4\pi)^2} \frac{19}{6}N_c }\,,
\end{align}
where for simplicity we have considered massless fermions ($\bar m_\text{f}=\bar m_\text{ad}=0$) and have approximated $\eta_{\rm ad}=\eta_{\rm f}=\eta_{\rm gh}\approx 0$. 

By performing a series expansion in the gauge coupling, we have
\begin{align}\label{eq: anomalous dimension 1 loop series expansion}
	\eta_g &= -\frac{\tilde g^2}{(4 \pi)^2}\left(  \frac{N_c}{3} \left(22-1\right)-\frac{4}{3}N_{\rm f}-\frac{8}{3}N_c N_{\rm ad}\right) +{\cal O }(\tilde{g}^4)\,,
\end{align}
where now the various matter contributions can easily be identified and compared to the well-known four dimensional structure of non-Abelian gauge theories. The shown anomalous dimension at the one-loop level in Eq.~\eqref{eq: anomalous dimension 1 loop series expansion} or the numerator of Eq.~\eqref{eq: anomalous dimension resumed} coincides with the beta function for a $SU(N_c)$ gauge group in the presence of $N_c$ real scalar fields transforming in the adjoint representation and $N_{\rm f}$ fundamental and $N_{\rm ad}$ adjoint fermions. The scalar-mode contribution appears from the emergence of an additional mode in the fifth dimensional gauge field $A_5$. Note that fermionic fields do not induce additional modes as five-dimensional Dirac fields are considered. While the anomalous dimension $\eta_g$ has a correct one-loop structure, higher-order contributions disagree due to lack of inclusion of some diagrams in the background approximation. In particular, higher order fermionic effects are missing as can be seen from the fact that the denominator of Eq.~\eqref{eq: anomalous dimension resumed} depends neither on $N_\text{f}$ nor on $N_\text{ad}$. These contributions can partially be included when feeding back the anomalous dimensions in Eq.~\eqref{eq: anomalous dimension of etag} arising from the cut-lines in Eq.~\eqref{eq:diagramatic flow}.  In fact, these terms are relevant in the $g\gg 0 $ limit where the contributions from the anomalous dimensions are not negligible and cannot be dropped. 

\begin{figure}
	\centering
	\includegraphics[width=.2\textwidth]{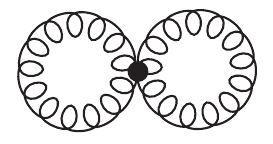} 
	\hspace{4ex}
	\includegraphics[width=.12\textwidth]{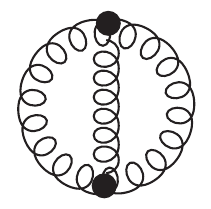} 
		\caption{Examples of two-loop diagrams. The left-hand side diagram produces the two-loop effect proportional to $[J(\bar R,0)]^2$, where $J(\bar R,0)$ is defined in Eq.~\eqref{eq: threshold function J}, while the right-hand side one gives a new type of threshold function $I(\bar R,0)$ different from $J(\bar R,0)$.}
	\label{Fig:two_loop}
\end{figure}

Furthermore, we comment on the $\bar R$ dependence in the resumed anomalous dimension~\eqref{eq: anomalous dimension resumed}. No compactification radius dependence appears explicitly  after the redefinition of the gauge coupling. However, this is a special case in which the anomalous dimensions contain only one specific class of loop diagrams which yield the same form of the threshold function $J(\bar R,0)$ as in Eq.~\eqref{eq: threshold function J}. See the left-hand side diagram in Fig.~\ref{Fig:two_loop}. Consequently, the flow equation for the gauge coupling takes a form as
\begin{align}
\p_t \left(\frac{Z^{-1}_kg^2}{2\pi  R} \right)&=   \bigg( c_1 J(\bar R,0) \left(\frac{Z^{-1}_kg^2}{2\pi  R}\right) \notag\\
&\hspace{-.67cm}+ c_2 [J(\bar R,0)]^2  \left(\frac{Z^{-1}_kg^2}{2\pi  R}\right)^2 +\cdots \bigg)  \frac{Z^{-1}_kg^2}{2\pi  R} \,,
\end{align}
where $c_i$ depend only on $N_f$ and $N_c$.  The redefinition of the gauge coupling \eqref{eq: dimensionless gauge coupling} leads to
\begin{align}
\p_t \tilde g^2 =   \Big[ (d_\text{eff}(\bar R)-4)  + \left(c_1 \tilde g^2 + c_2\tilde g^4 +\cdots \right)\Big]\tilde g^2\,.
\end{align}
Here, the term in the second parentheses in square brackets corresponds to the anomalous dimension \eqref{eq: anomalous dimension resumed} free from the compactification radius.
However, in more general case beyond the one-loop resummation such as the right-hand side diagram in Fig.~\ref{Fig:two_loop}, explicit $\bar R$ dependencies appear that are not absorbed by the definition of $\tilde g$. These cannot be appreciated in the background anomalous dimension given that only accounts for one class of diagrams with the same $\bar R$ dependence properly absorbed in the definition of $\tilde g$. More specifically, the flow equation reads
\begin{align}
\p_t \left(\frac{Z^{-1}_kg^2}{2\pi  R} \right) &=  \bigg[ c_1 J(\bar R,0)\left( \frac{Z^{-1}_kg^2}{2\pi  R}\right) \nn
&\hspace{-.5cm}  + \bigg( c_2 [J(\bar R,0)]^2  + d_2 I(\bar R,0) +\cdots  \bigg)\left(\frac{Z^{-1}_kg^2}{2\pi  R}\right)^2  \nn
&\hspace{0cm}+\cdots \bigg]\left(\frac{Z^{-1}_kg^2}{2\pi  R}\right)\,,
\end{align}
for which after the redefinition of the gauge coupling \eqref{eq: dimensionless gauge coupling}, we have
\begin{align}
&\p_t \tilde g^2 = \Big[ (d_{\rm eff} (\bar R) -4) \nn
&+  \left\{ c_1 \tilde g^2 + \left( c_2 + d_2\tfrac{ I(\bar R,0)}{J(\bar R,0)}   +\cdots \right)\tilde g^4 +\cdots \right\}\Big]\tilde g^2\,,
\end{align}
where  a new threshold function $I(\bar R,0)$ with a factor $d_2$ has been introduced. $\bar R$ dependences such as ${I(\bar R,0)}/{J(\bar R,0)}$ remain. In the following section, we discuss how the resumed computation can be improved by adopting $\overline{\rm MS}$ results. 
%%%%%%%%%%%%%%%
%%%%%%%%%%%%%%%
%%%%%%%%%%%%%%%
%%%%%%%%%%%%%%%

\subsection{UV \& IR fixed points}

%%%%%%%%%%%%%%%%%%%%%%%%%%%%
%%%%%%%%%%%%%%%%%%%%%%%%%%%%
\begin{figure*}
	\includegraphics[width=.95\textwidth]{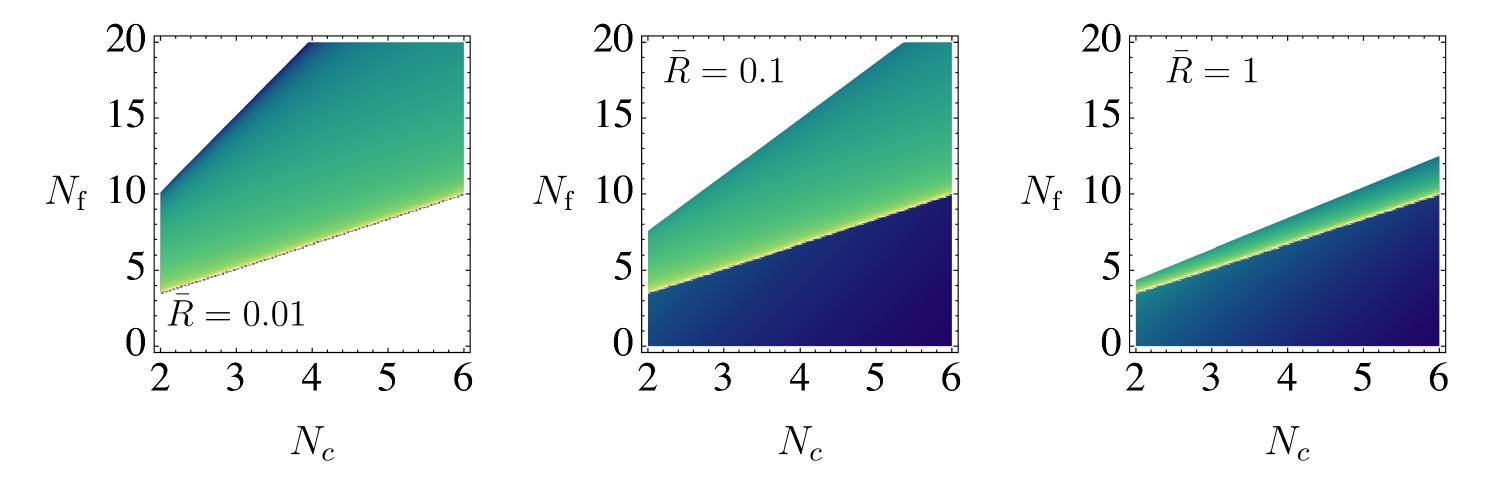}	\vspace{-.75cm}
	
	\hspace{0.00001ex}
	
	\includegraphics[width=.95\textwidth]{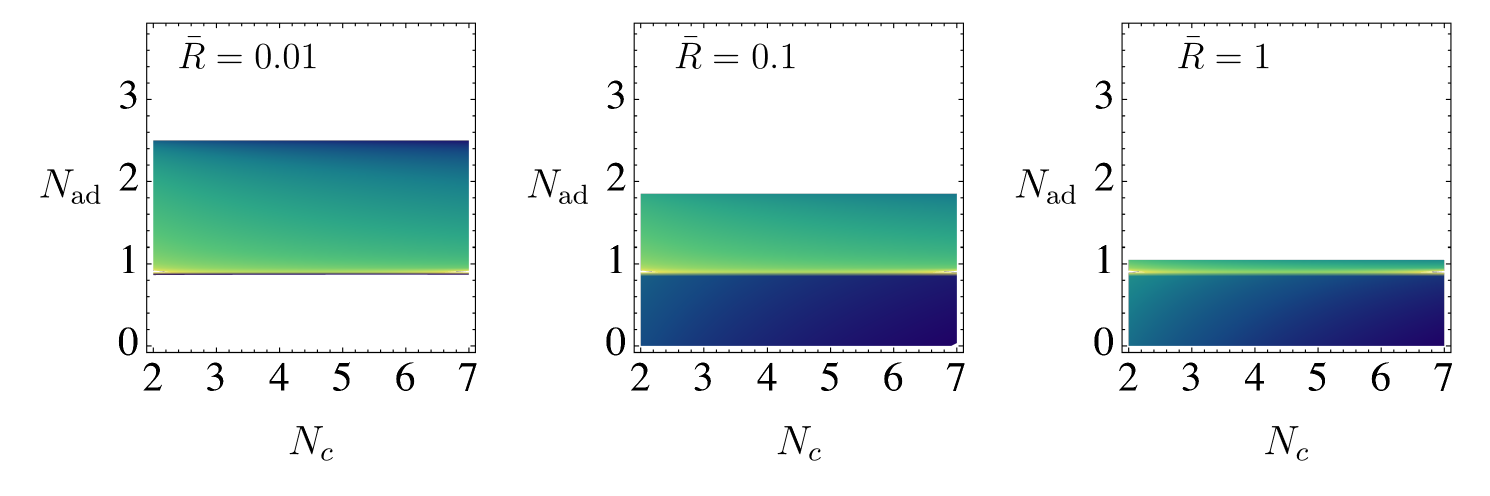}	\vspace{-.5cm}
	
	\caption{In the top (bottom) row, the non-trivial UV and IR fixed-point solutions for an $SU(N_c)$ gauge coupling in the presence of $N_{\rm f}$ ($N_{\rm ad}$) fermion fields transforming in the fundamental (adjoint) representations and at different compactification radii: $\bar R=0.01$ (left column), $\bar R=0.1$ (centre column) and $\bar R=1$ (right column). Brighter colours indicate a larger magnitude of the fixed point. In order to account for beyond one-loop fermionic contributions, we have employed three-loop $\overline{\rm MS}$ results in the anomalous dimension $\eta_g$ in Eq.~\eqref{eq: beta function gtilde}. }
	\label{Fig: UV fixed points regimes in change of R}
\end{figure*}
%%%%%%%%%%%%%%%%%%%%%%%%%%%%
%%%%%%%%%%%%%%%%%%%%%%%%%%%%

Let us study the fixed-point structure of the gauge coupling using the flow equation \eqref{eq: beta function gtilde} and as a function of the compactification radius. Yang-Mills theories coupled to fermions are known to feature a very rich phase structure in the 4-dimensional limit ($\bar R\to 0$). Depending on the number of fermions and the representation under which they transform, we can find different fixed-point solutions. 
For example, in the QCD-like scenario with small number of fundamental fermions, the theory has a trivial UV fixed point ($\tilde g_*=0$)  associated to asymptotic freedom~\cite{Gross:1973id,Politzer:1973fx}  and in the IR regime the system turns strongly coupled. Increasing the number of fermions from the QCD scenario, non-trivial IR solutions appear in the beta function. These are known as CBZ fixed points and are roots of the beta function caused by the cancellations of different loop orders. Therefore, contributions beyond one loop are determinant in this limit. Furthermore, in the many fermion limit, asymptotic freedom is lost leading to a UV Landau pole and the non-trivial IR fixed point evolves into a trivial (Gaussian) solution. 

Here, we analyse the fixed-point solutions as a function of the effective dimensionality parametrised by the dimensionless compactification radius $\bar R$. For simplicity, we consider the massless fermion limit and with the periodic boundary conditions $\beta_\text{f}=\beta_\text{ad}=0$. The presence of masses simply triggers the decoupling of fermionic degrees of freedom and hence the UV and IR fixed points are not significantly affected by such parameters. However, masses, boundary conditions or a non-trivial vacuum structure will leave imprints in the integrated flows at finite $\bar R$ and $k$, see Ref.~\cite{Pastor-Gutierrez:2022rac}. We denote $g^2_*(\bar R, N_\text{f},N_\text{ad},N_c)$ as the non-trivial fixed-point solutions for arbitrary compactification radius, number of fermionic fields and colours. These solutions are found in Eq.~\eqref{eq: beta function gtilde} by solving $d_\text{eff}(\bar R)-4 + \eta_g=0$,
 \begin{align}
	\tilde g^2_*(\bar R, N_\text{f},N_\text{ad},N_c)= \frac{(d_\text{eff}(\bar R) -4)(4\pi)^2}{b_0+c(d_\text{eff}(\bar R) -4)}\,,
	\label{eq: fixed point solution}
\end{align}
where $b_0=7N_c - \frac{4}{3} N_{\rm f} - \frac{8}{3}N_{\rm ad}N_c$ and $c= \frac{19}{6}N_c$. For $\bar R=0$ ($D=4$), the fixed-point solution \eqref{eq: fixed point solution} vanishes leading to  the trivial UV fixed point characterizing asymptotic freedom. As the compactification radius increases, $g^2_*$ takes a finite value. This fact provides evidence for the non-perturbatively renormalizability, in sense of asymptotic safety~\cite{Hawking:1979ig,Nagy:2012ef}, of the theory. See the discussion at the end of this section.

From Eq.~\eqref{eq: beta function gtilde} we find that in the absence of threshold effects, the dependence on the compactification radius only enters via the effective dimensionality $d_\text{eff}(\bar R)$. As stated in the end of the previous subsection, this follows from the background approximation only accounting for one specific family of diagrams producing $J(\bar R,0)$ whose $\bar R $ dependence is fully absorbed in the definition of $\tilde g$.  Moreover, neglecting these dependencies permits us to conveniently improve Eq.~\eqref{eq: fixed point solution} by implementing $\eta_g$'s  from computations accounting for the complete higher order fermionic contributions lacking in the background approximation. Furthermore, we will employ three-loop $\overline{\rm MS}$ results~\cite{vanRitbergen:1997va,Vermaseren:1997fq,Pica:2010xq} for $\eta_g$ instead of Eq.~\eqref{eq: anomalous dimension resumed} for the following study. This approximation implicitly assumes the ratio ${I(\bar R,0)}/{J(\bar R,0)}\approx1$ and hence the $\bar R$ dependence of the diagrams not accounted in the background approximation to be small. 

Before proceeding with the analysis, let us first investigate the appearance of fixed point solutions entertaining a perturbative expansion of the anomalous dimension in Eq.~\eqref{eq: beta function gtilde},
\begin{align}
4-d_\text{eff}(\bar R) = \eta_g&= 	\tilde g^2_*(\bar R) \left[ f^{(1)} + \tilde g^2_*(\bar R)f^{(2)}(\bar R) + \cdots\right]\,.
	\label{eq: perturbative analyisis of beta function}
\end{align}	
Here, $f^{(i)}$ are the $i$-order loop contributions dependent on the gauge group and fermion content. While due to the redefinition of the gauge coupling \eqref{eq: dimensionless gauge coupling}, the full $R$-dependence of the one-loop coefficient $f^{(1)}$ is translated into the effective dimension \eqref{eq: effective dimension}, for higher-loop terms a residual contribution remains in the $f^{(i>1)}$  functions given the non-trivial powers of 
the threshold function \eqref{eq: threshold function J}.
In $D=4$ ($R\to0$), we find two solutions vanishing the right-hand side of Eq.~\eqref{eq: perturbative analyisis of beta function}:
\begin{align}
%R\to0 :	
\begin{cases}
	\tilde g^2_*(0)=	0 &({\rm Gaussian})\,,\\
	\\
	 f^{(1)} + \tilde g^2_*(0)f^{(2)}(0) + \cdots=0 &({\rm CBZ})\,.
	\end{cases}       
\end{align}
%per
While the Gaussian fixed point is a UV fixed point and characterises asymptotic freedom, the non-trivial CBZ fixed point is an IR solution given as a root of the several loop contributions in polynomial of $g^2$. 

On the other hand, in the  $D=5$ ($R\to \infty$) limit, the non-trivial fixed-point solution is obtained when
\begin{align}
\tilde g^2_*(\infty) f^{(1)} + \tilde g^4_*(\infty)f^{(2)}(\infty) + \cdots =4-d_\text{eff}(\infty)=-1 \,,
\end{align}	
is satisfied. In this limit, the roots of the beta function cannot be only linked to the roots of the anomalous dimension and hence the new fixed points cannot be related to the four-dimensional CBZ solution.

%%%%%%%%%%%%%%%%%%%%%%%%%%%%
%%%%%%%%%%%%%%%%%%%%%%%%%%%%
\begin{figure*}[th!]
	\includegraphics[width=1\textwidth]{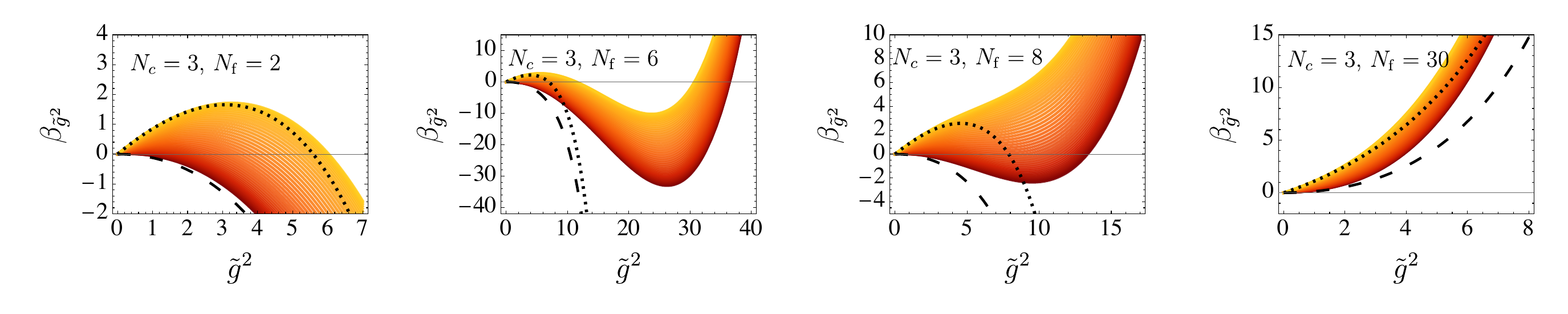}	\vspace{-.75cm}
	
	\includegraphics[width=1.\textwidth]{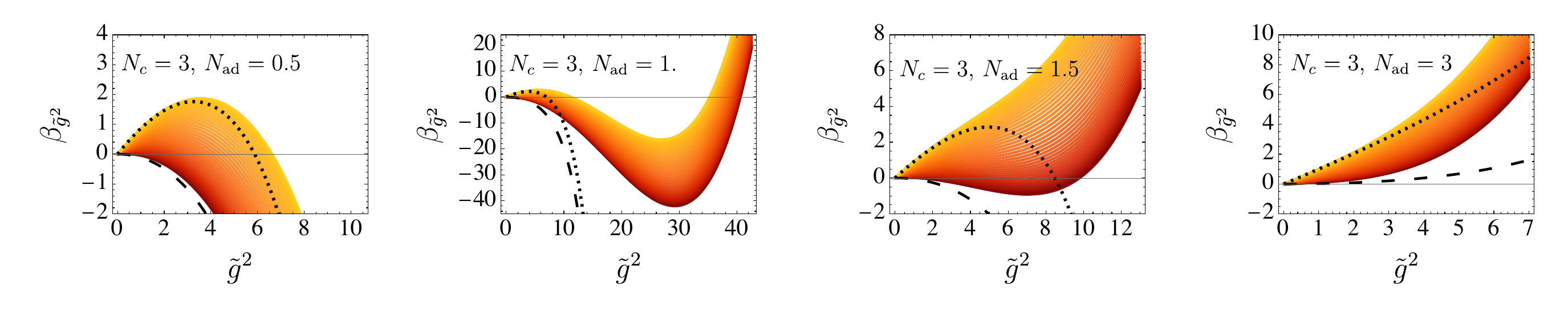}		\vspace{-1.cm}
	\caption{
	$SU(3)$ gauge coupling beta functions as a function of the coupling for different fermion content in the fundamental (top row) and adjoint (bottom row) representations. As plain lines from red (dark) to yellow (bright), we show the evolution from $\bar R=0$ to $\bar R\to \infty$ employing the three-loop $\overline{\rm MS}$ results. In black we show the resumed beta functions at $\bar R=0$ (black dashed) and $\bar R\to \infty$ (black dotted). The resumed anomalous dimension does not show the CBZ fixed point in the $\bar R=0$ as lacks the complete two-loop contributions. Nevertheless, in the large and few $N_{\rm f},\, N_{\rm ad}$ limits, both beta functions qualitatively agree. Non-integer number of fermions are employed in some instances for best display of qualitative features.}
	\label{Fig: beta function as a function of g2}
\end{figure*}
%%%%%%%%%%%%%%%%%%%%%%%%%%%%
%%%%%%%%%%%%%%%%%%%%%%%%%%%%

%%%%%%%%%%%%%%%%%%%%%%%%%%%%
%%%%%%%%%%%%%%%%%%%%%%%%%%%%
\begin{figure*}[th!]
	\includegraphics[width=.3\textwidth]{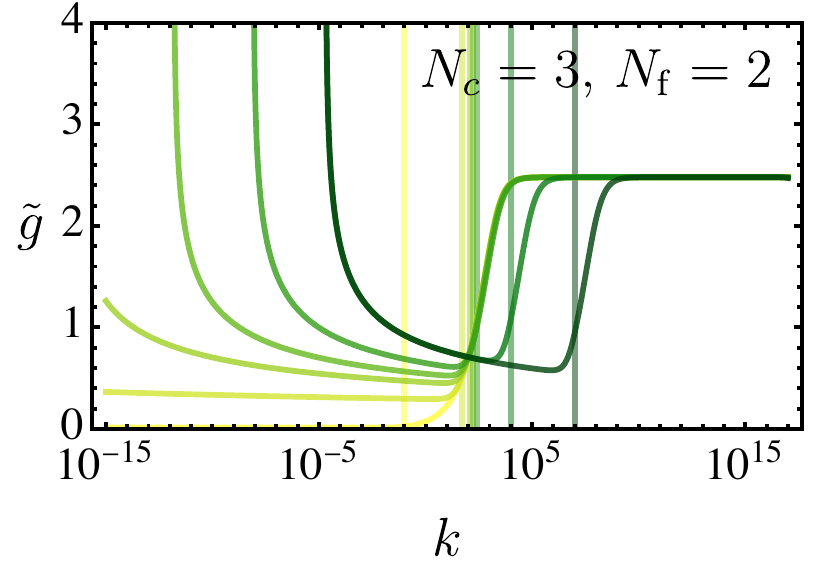}\hspace{.75cm}	\includegraphics[width=.3\textwidth]{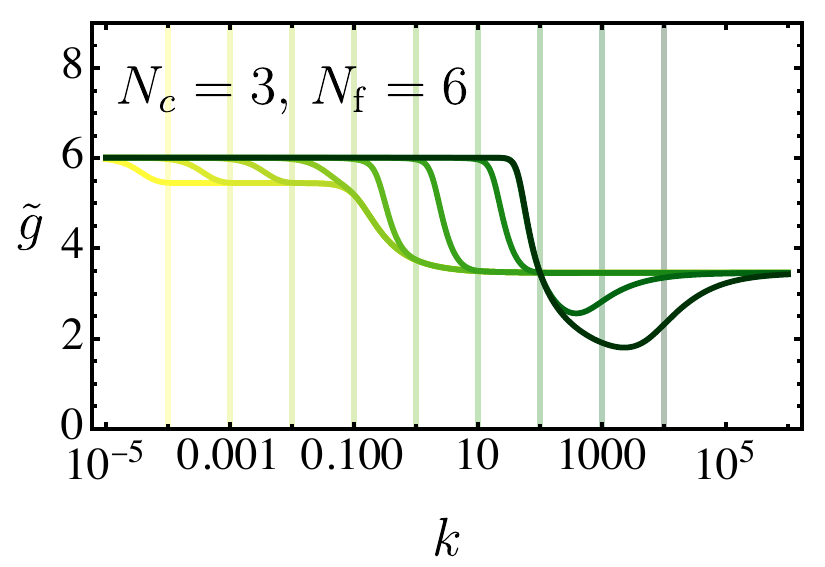}\hspace{.75cm}\includegraphics[width=.3\textwidth]{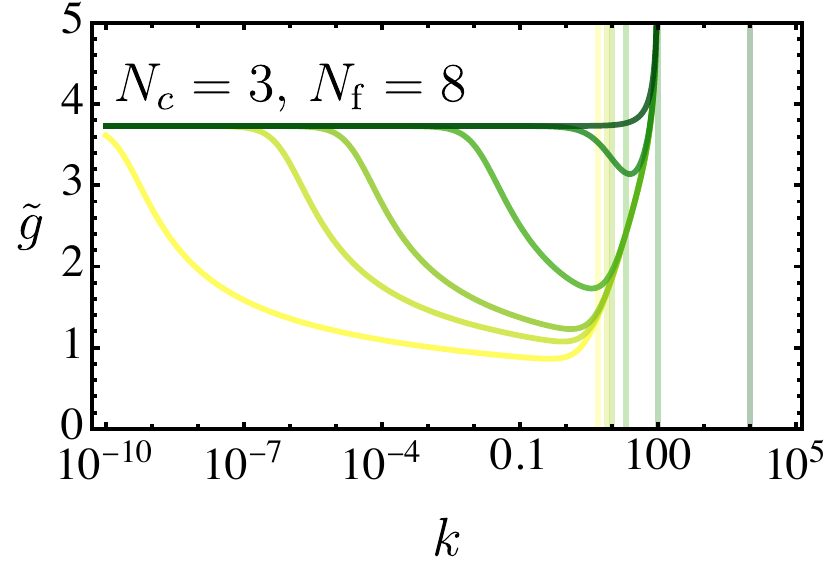}	
	\caption{Integrated renormalization group flows of the $SU(3)$ coupling for different fermion content. From left to right, we increase the number of fermions  $N_{\rm f}=2,\, 6,\, 8$ transforming in the fundamental representation, while $N_\text{ad}=0$. In each plot we show different curves for which the onset of the extra-dimension triggers at different scales. The vertical lines denote the onset of the extra-dimension where $R/k=1$ is satisfied. 
	}
	\label{Fig: flow of gauge coupling}
\end{figure*}
%%%%%%%%%%%%%%%%%%%%%%%%%%%%
%%%%%%%%%%%%%%%%%%%%%%%%%%%%

In Fig.~\ref{Fig: UV fixed points regimes in change of R}, we display the regions on the $N_c$--$N_{\rm f}/N_{\rm ad}$ planes where the beta function for the gauge coupling \eqref{eq: beta function gtilde} shows together both non-trivial UV and IR fixed points. In the $\bar R\to 0$ limit, we find the non-trivial IR fixed-point solutions for $N_{\rm f}$ and $N_{\rm ad}$ which satisfy $\eta_g=0$. This region is known as the CBZ window and their properties such as boundaries and magnitude of the fixed point have been widely studied with perturbative~\cite{Banks:1981nn,Dietrich:2006cm,Pica:2010xq,Ryttov:2010iz,Ryttov:2016hal,Ryttov:2016ner,Ryttov:2017kmx,Ryttov:2018uue,Gracey:2023unc}, and non-perturbative methods~\cite{Braun:2006jd,Gies:2005as,Hopfer:2014zna} as well as lattice simulations~\cite{Ryttov:2009yw, Ryttov:2007sr,LatticeStrongDynamics:2018hun,LatticeStrongDynamics:2020uwo,Hasenfratz:2019dpr,Hasenfratz:2022zsa,Hasenfratz:2023wbr}. For a larger number of fermions, the non-trivial IR fixed point turns into a Gaussian fixed point, loosing asymptotic freedom in the UV limit. It is important to remind about a difference with the standard gauge-fermion CBZ window picture, in the current $D=5$ scenario there are additional $N_c$  scalar degrees of freedom in the $\bar R \to 0$ limit associated to the five dimensional modes of the gauge field. Their contribution reflects in a subtle fewer number of fermionic degrees of freedom necessary to reach the conformal window.

As the compactification radius is increased, the growth of dimensionality causes the canonical dimension of the gauge coupling to increment turning the marginal gauge coupling into a relevant one. Moreover, as was shown in Ref.~\cite{Gies:2003ic,Pastor-Gutierrez:2022rac}, pure Yang-Mills theories displaying asymptotic freedom in the UV limit evolve to asymptotic safety.  This is evidently seen in the appearance of a non-trivial fixed point in the lower boundary of the CBZ window and below.  The increment of the canonical dimension of the gauge coupling contributes positively to the flow alike the addition of fermionic matter. Hence, increasing the dimensionality will require a lower number of fermions to trigger a singular solution. This causes the CBZ window  to shrink with the increment of $\bar R$. However, we note a remanent of the CBZ fixed point at $\bar R\to \infty$ in the lower boundary of the window (bright yellow band), leading to the simultaneous presence of IR and UV fixed-point solutions. This will be discussed in detail below. 

The dependence of the beta function on the change of dimension is best studied as a function of the gauge coupling. This is shown in Fig.~\ref{Fig: beta function as a function of g2}. Here, both of the approximations employed for $\eta_g$, the resumed shown in Eq.~\eqref{eq: anomalous dimension resumed} (black lines) and the three-loop perturbative (red-orange lines), are shown for different $\bar R $. While both computations agree well at small $\tilde g^2$, the resumed anomalous dimension lacks to reproduce the non-trivial CBZ fixed point as does not account for the complete two loop and higher structure. 

As shown in Ref.~\cite{Pastor-Gutierrez:2022rac}, in the pure Yang-Mills or QCD-like limit, the beta function experiences a relevant transformation as the gauge coupling's dimension changes with the compactification radius. In the left most panels of  of Fig.~\ref{Fig: beta function as a function of g2}, it can be seen how the beta function with a single Gaussian UV fixed point develops an additional non-trivial solution. In other words, the change in dimensionality turns the Gaussian UV fixed point into an IR fixed-point solution and generates a non-trivial UV fixed point.

Furthermore, configurations which in $D=4$ show a CBZ fixed point can evolve to two different scenarios. In the upper boundary of the CBZ window, where $g^2_*(\bar R= 0, N_\text{f},N_\text{ad},N_c)$, the non-trivial IR fixed point is lost leading to a five-dimensional theory with a trivial IR and a UV Landau pole (third plot in each row in Fig.~\ref{Fig: beta function as a function of g2}). In the second case, we find that the IR CBZ fixed point does not disappear in the $\bar R \to \infty$ limit and as the new non-trivial UV fixed point emerges. This leads to three solutions being present and  to UV and IR finite trajectories to be realisable (second plot in each row of Fig.~\ref{Fig: beta function as a function of g2}). 

The many flavour limit is unaffected by the change of dimensionality. This can be seen in the last panels of each row in which the five-dimensional beta function does not present any new solutions.
 
In Fig.~\ref{Fig: flow of gauge coupling}, we depict different trajectories of the integrated gauge coupling flow as a function of the cut-off scale and considering various on-sets of the extra-dimension. For a fixed number of colours $N_c=3$ and  $N_{\rm f}=2$, the inter-dimensional flows connect a four dimensional strong IR QCD-like scenario with a UV finite five-dimensional setting as shown in the left-most panel of Fig.~\ref{Fig: flow of gauge coupling}. Here, from bright yellow to dark green the on-set scale of the extra-dimension is delayed with respect to the cut-off scales $k$. Depicted by vertical lines with the respective colour of the trajectory, we show the scale at which $R/k=1$ and hence, the onset of the extra dimension.  

For $N_{\rm f}=6$ we find a particularly interesting scenario (middle panel in Fig.~\ref{Fig: flow of gauge coupling}). From the UV, we commence the flow at approximately the non-trivial fixed point in the $D=5$ limit. If the system is kept in the $\bar R\to \infty$ limit, trajectories evolve towards either a Gaussian or interacting IR fixed point. The former can be appreciated in the darkest trajectories and the latter in the brightest. However, as soon as the system senses the $D=4$ limit, both IR solutions are attracted towards the CBZ fixed point .

The right-most plot in Fig.~\ref{Fig: flow of gauge coupling} shows trajectories for $N_{\rm f}=8$ in which the CBZ solution exists in the $D=4$ limit but no fixed point is present in $D=5$.  

Finally,  we discuss the critical exponent of the gauge coupling denoted here by $\nu_{\tilde g^2}$. At a fixed point $\tilde g_*^2$, the gauge coupling behaves as a power law
\begin{align}
\tilde g^2 \sim k^{-\nu^{-1}_{\tilde g^2}}\,.
\end{align}
From the beta function of the gauge coupling \eqref{eq: beta function gtilde}, the critical exponent is given by
\begin{align}
\nu_{\tilde g^2}^{-1} &= - \frac{\p \beta_{\tilde g^2}}{\p \tilde g^2}\bigg|_{\tilde g^2=\tilde g^2_*}\nn
&=- (d_{\rm eff}(\bar R) - 4 + \eta_g)|_{\tilde g^2=\tilde g^2_*} - \frac{\p \eta_g}{\p \tilde g^2}\bigg|_{\tilde g^2=\tilde g^2_*} \tilde g^2_*\,.
\label{eq: critical exponent of g2}
\end{align}
At the Gaussian fixed point $\tilde g^2_*=0$ for which $\eta_g=0$, one obtains $\nu_{\tilde g^2}^{-1} = 4- d_{\rm eff}(\bar R)$. Since $d_{\rm eff}(\bar R)>4$ for finite $\bar R$ as shown in Fig.~\ref{Fig:effective dimension},  we observe $\nu_{\tilde g^2}^{-1} <0$ and therefore the gauge coupling at the Gaussian fixed point is an irrelevant coupling in extra-dimensional spacetimes. In other words, the Gaussian fixed point is an IR attractive fixed point. The gauge coupling becomes marginal only when $d_{\rm eff}(\bar R=0)=4$. As is well known, Yang-Mills theories in $D=4$ exhibit the marginally relevant behaviour of the gauge coupling if the fermion numbers do not exceed a certain threshold. This fact can be seen from $\eta_g<0$ for a small perturbation from the Gaussian fixed point as $0<g_*^2\ll 1$. In such a case, one has $\nu_{\tilde g^2}^{-1}\ll 1$ so that 
\begin{align}
\tilde g^2 \sim  e^{ -\log k/\nu_{\tilde g^2}} \sim  \nu_{\tilde g^2}^{-1} \log k\,.
\end{align}
This logarithmic running is a natural result inferred from the perturbative computation.

On the other hand, at the non-trivial fixed points $\tilde g^2_*\neq 0$, the first term on the right-hand side in Eq.~\eqref{eq: critical exponent of g2} vanishes, so that the critical exponent reads
\begin{align}
\nu_{\tilde g^2}^{-1} &=- \frac{\p \eta_g}{\p \tilde g^2}\bigg|_{\tilde g^2=\tilde g^2_*} \tilde g^2_*\,.
\end{align}
For the anomalous dimension \eqref{eq: anomalous dimension resumed}, there is the non-trivial UV fixed point \eqref{eq: fixed point solution} at which the critical exponent is found to be
\begin{align}
\nu_{\tilde g^2}^{-1}  = \frac{\left(b_0+c(d_{\rm eff}(\bar R) -4) \right)\left(d_{\rm eff}(\bar R) -4 \right)}{b_0}\,.
\end{align}
For $\bar R>0$, we see that $\nu_{\tilde g^2}^{-1} >0$ for $b_0+c(d_{\rm eff}(\bar R) -4)>0$, namely the gauge coupling is relevant. For large fermion numbers, $b_0+c(d_{\rm eff}(\bar R) -4)<0$, leading to the fixed point \eqref{eq: fixed point solution} turning positive. Hence, for positive values of the non-trivial UV fixed point, the corresponding critical exponent always takes positive values, implying a relevant direction.

The existence of such a non-trivial UV fixed point provides evidence for extra-dimensional Yang-Mills theories to be non-perturbatively renormalisable in terms of asymptotic safety. However, the current setup for the truncated effective action cannot determine the number of relevant couplings and hence, the non-perturbative renormalisability of the theory cannot be concluded. In order to determine the dimensionality of the UV critical surface, higher dimensional operators such as $(F_{\mu\nu}F^{\mu\nu})^n$ have to be taken into account. This project will be reported elsewhere.

%%%%%%%%%%%%%%%
%%%%%%%%%%%%%%%
\begin{figure*}[ht!]
	\begin{center}
		\begin{minipage}{0.45\textwidth}
			\begin{tabular}{| c|| c| c| c |c |c }
				\toprule
				\makebox[1.6cm]{$\mathcal H_\text{sym}/\text{Phase}$} &  \makebox[1cm]{config.} & \makebox[1.1cm]{$\theta_1$}  & \makebox[1.1cm]{$\theta_2$}  &   \makebox[1.1cm]{$\theta_3$}  \\[1ex]  \hline\hline
				%			\midrule
				$SU(3)/\text{confined}$   & $X$ & --- & --- & --- \\[1ex] \hline
				& $A_1$ & $0$ & $0$ & $0$  \\[1ex]  
				$SU(3)/\text{deconfined}$  & $A_2$ & $\frac{2}{3}\pi$ & $\frac{2}{3}\pi$ & $\frac{2}{3}\pi$  \\[1ex]
				& $A_3$ &	$-\frac{2}{3}\pi$ & $-\frac{2}{3}\pi$ & $-\frac{2}{3}\pi$  \\ [1ex]\hline 
				%			\midrule
				& $B_1$  & $\alpha$   & $\alpha$ & $-2\alpha$  \\[1ex]
				$SU(2)\times U(1)/\text{split}$ & $B_2$ &  $\alpha$   & $-2\alpha$ & $\alpha$  \\[1ex]
				& $B_3$ &  $-2\alpha$   & $\alpha$ & $\alpha$  \\[1ex] \hline  
				%			\midrule
				& $C_1$ &   $\beta_1$ & $\beta_2$ & \scriptsize{$-\beta_1-\beta_2$} \\[1ex]
				$U(1)\times U(1)/\text{reconfined}$  &  $C_2$ &   $\beta_1$  & \scriptsize{$-\beta_1-\beta_2$}  & $\beta_2$ \\[1ex]
				& $C_3$ &  \scriptsize{$-\beta_1-\beta_2$} & $\beta_1$ & $\beta_2$ \\[1ex] \hline 
				%			\bottomrule
			\end{tabular}
		\end{minipage}
		\hspace{1.5cm}\begin{minipage}{0.45\textwidth}
			\vspace{-0.35cm}\includegraphics[width=.85\textwidth]{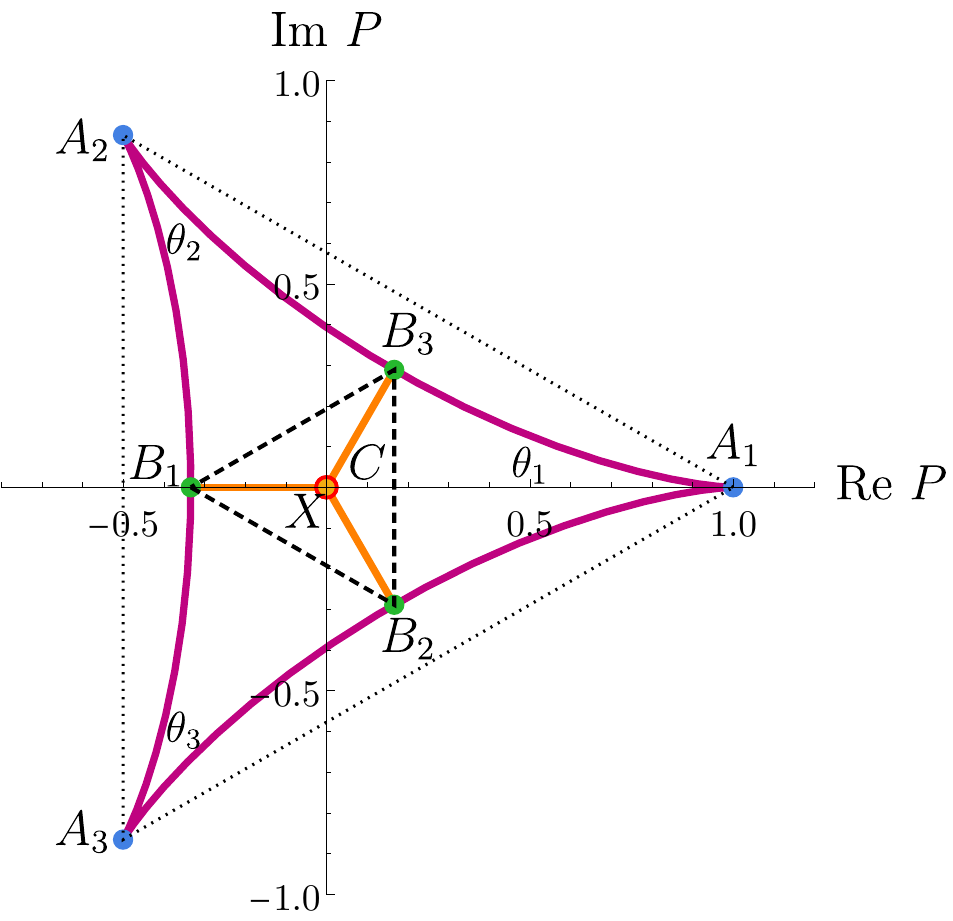}
		\end{minipage}
		
		\caption{
			\label{Table: summary of phases}
			Configurations of Aharonov-Bohm phases and breaking patterns of $SU(3)$ gauge symmetry, namely $SU(3)\to \mathcal H_{\rm sym}$ via the Hosotani mechanism~\cite{Cossu:2009sq,Cossu:2013ora}.  Here, $\alpha\neq0$, $\beta_1\neq \beta_2,\,-\frac{1}{2}\beta_2,\,-2\beta_2$.  The confined configuration $X$ arises from the Haar-measure distribution which is uniformly and randomly distributed so as to be $P=0$~\cite{Cossu:2013ora}. On the right hand side, the Polyakov loop configurations (defined in Eq.~\eqref{eq: Polyakov loop}) for the phases summarised on left hand side Table are displayed. 
		}
	\end{center}
\end{figure*}
%%%%%%%%%%%%%%%
%%%%%%%%%%%%%%%

\section{Gauge Potential and Aharonov-Bohm phases}
\label{sec: Potential of AB phases}

In this section, we study the phase structure of the four- and five-dimensional gauge effective. To this end, we consider the effective potential for the AB phases and the vacuum structure in the presence of fermions. The relevant flow equation for the effective potential can be obtained employing heat-kernel techniques and reads
\begin{align}
	&\p_t V(\theta_H) = \nn[1ex]
	&\,+\frac{5k^4}{2(4\pi)^2}\left( 1-\frac{\eta_g}{6} \right)\sum_{i,j=1}^{N_c} h(\bar R;\theta_i-\theta_j;0)  \nn
	&\,-\frac{k^4}{(4\pi)^2}\left( 1-\frac{\eta_\text{gh}}{6} \right)\sum_{i,j=1}^{N_c} h(\bar R;\theta_i-\theta_j;0)\nn
	&\,-\frac{2N_{\rm ad}k^4}{(4\pi)^2}\left( 1-\frac{\eta_\text{ad}}{5} \right)\sum_{i,j=1}^{N_c} h(\bar R;\theta_i-\theta_j-\beta_{\rm ad};\bar m_\text{ad})\nn
	&\,-\frac{2N_{\rm f}k^4}{(4\pi)^2}\left( 1-\frac{\eta_\text{f}}{5} \right)\sum_{i=1}^{N_c} h(\bar R;\theta_i-\beta_{\rm f};\bar m_\text{f})\,.\hspace{-.0cm}
	\label{eq: Flow of the potential}
\end{align}
Here, the threshold function with a mass and AB phases is given by 
\begin{align}
	\hspace{-.25cm}h(\bar R; \theta_H; \bar m)	&= \frac{i \pi  \bar R}{2 \sqrt{1+\bar m^2}}\bigg[ \cot \left(\frac{1}{2} \bar R \left(\theta_H+2 i \pi\sqrt{1+\bar m^2}\right)\right)\nn[1ex]
	&\hspace{-.1cm}+\cot \left(-\frac{1}{2} \bar R \left(\theta_H- 2 i \pi \sqrt{1+\bar m^2}\right)\right) \bigg]\,.
\label{eq: threshold function with m and theta}
\end{align}
Note that setting $\theta_H=0$ in Eq.~\eqref{eq: threshold function with m and theta} reduces to $J(\bar R; \bar m)$ defined in Eq.~\eqref{eq: threshold function J}, i.e. $h(\bar R;0;\bar m)=J(\bar R;\bar m)$. For a detailed derivation we refer to Appendices of Ref.~\cite{Pastor-Gutierrez:2022rac}.

A key quantity for understanding the gauge phase structure is the Polyakov loop which is defined as $P = \tr W/N_c$. Here, $W$ denotes the Wilson line along $S^1$ compactified direction (see e.g. Ref.~\cite{Pastor-Gutierrez:2022rac} for its explicit definition). The Polyakov loop is the gauge independent order parameter for distinguishing the confined ($P=0$) and deconfined ($P\neq 0$) phases. Especially $P\neq 0$ implies that the global $Z_{N_c}$ centre symmetry of $SU(N_c)$ is spontaneously broken. While in $SU(N_c)$ gauge theories with adjoint matter the $Z_{N_c}$ centre symmetry is preserved, the introduction of fundamental fermions leads to the explicit $Z_{N_c}$ centre symmetry breaking. As a result, if no other matter fields are included, the minimum is always in the deconfined phase \cite{Pastor-Gutierrez:2022rac}.

As a specific example, we consider $N_c=3$ for which there are three AB phases ($\theta_1$, $\theta_2$, $\theta_3$), but only two phases among them are independent parameters thanks to the traceless condition $\theta_1+\theta_2+\theta_3=0$. Depending on the configurations of AB phases,  $SU(3)$ gauge symmetry is broken into different subgroups $\mathcal H_\text{sym}$. The breaking patterns for possible configurations of AB phases are listed in Fig.~\ref{Table: summary of phases}. Moreover, for these AB phases, we show the values of the Polyakov loop defined as
\begin{align}\label{eq: Polyakov loop}
P=\frac{1}{3}(e^{i\theta_1} + e^{i\theta_2} + e^{i\theta_3})\,,
\end{align}
on the right-hand side panel of Fig.~\ref{Table: summary of phases}. 

%%%%%%%%%%%%%%%
%%%%%%%%%%%%%%%
\begin{figure*}
\hspace{-0.2cm}	\includegraphics[width=0.41\textwidth]{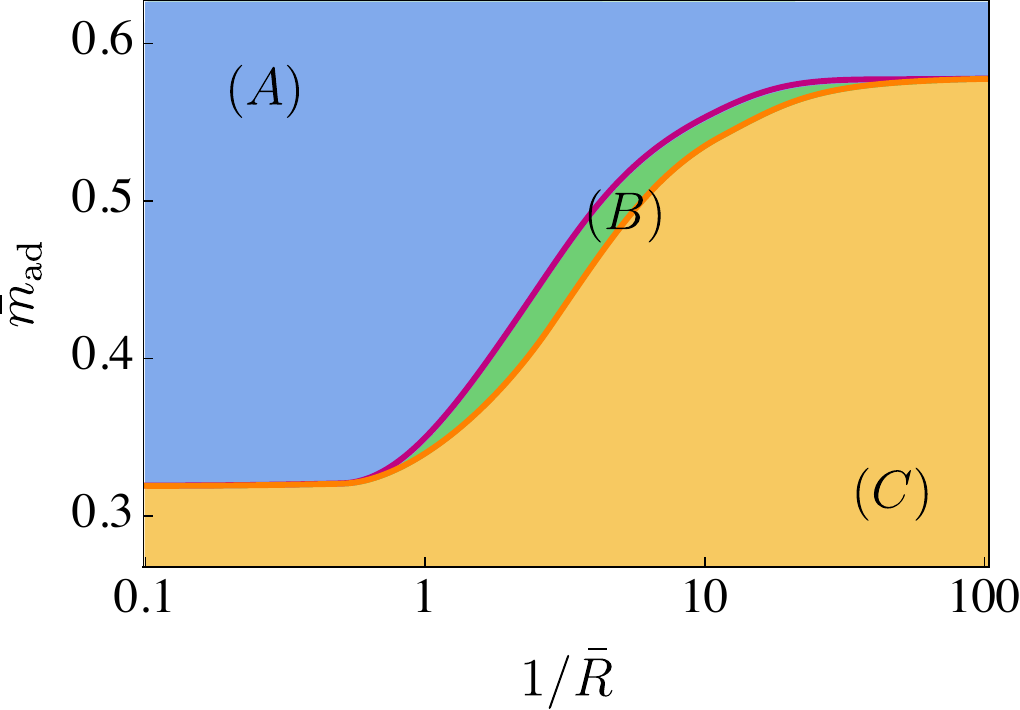}\hspace{1cm}
	\includegraphics[width=0.25\textwidth]{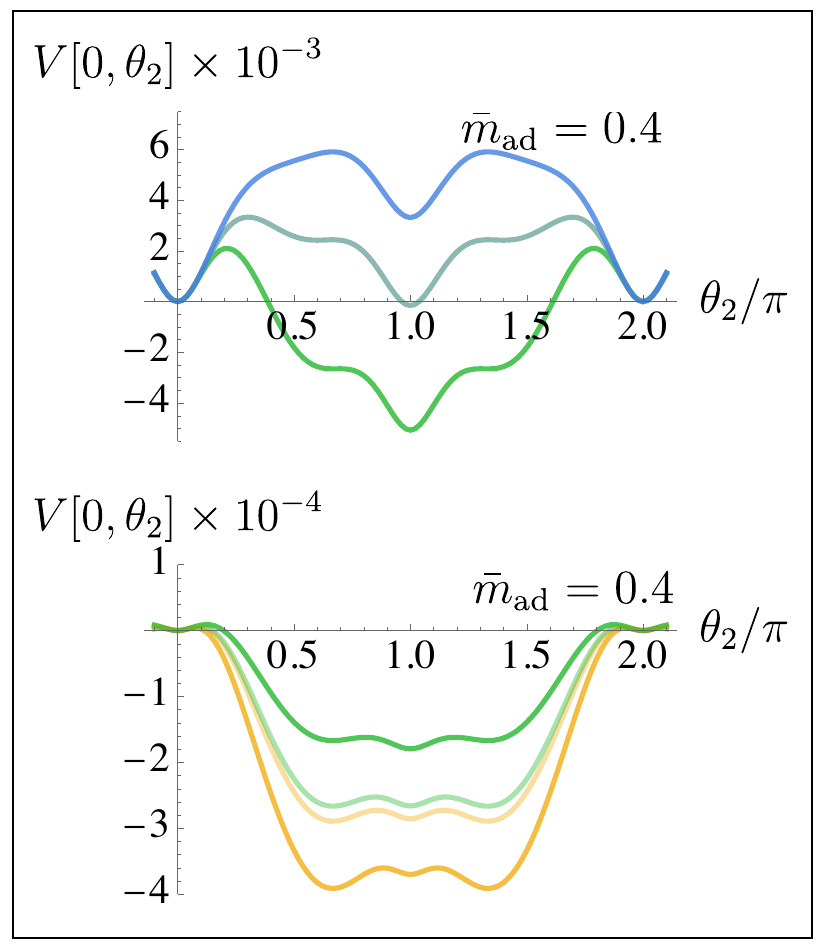} 
	\includegraphics[width=0.25\textwidth]{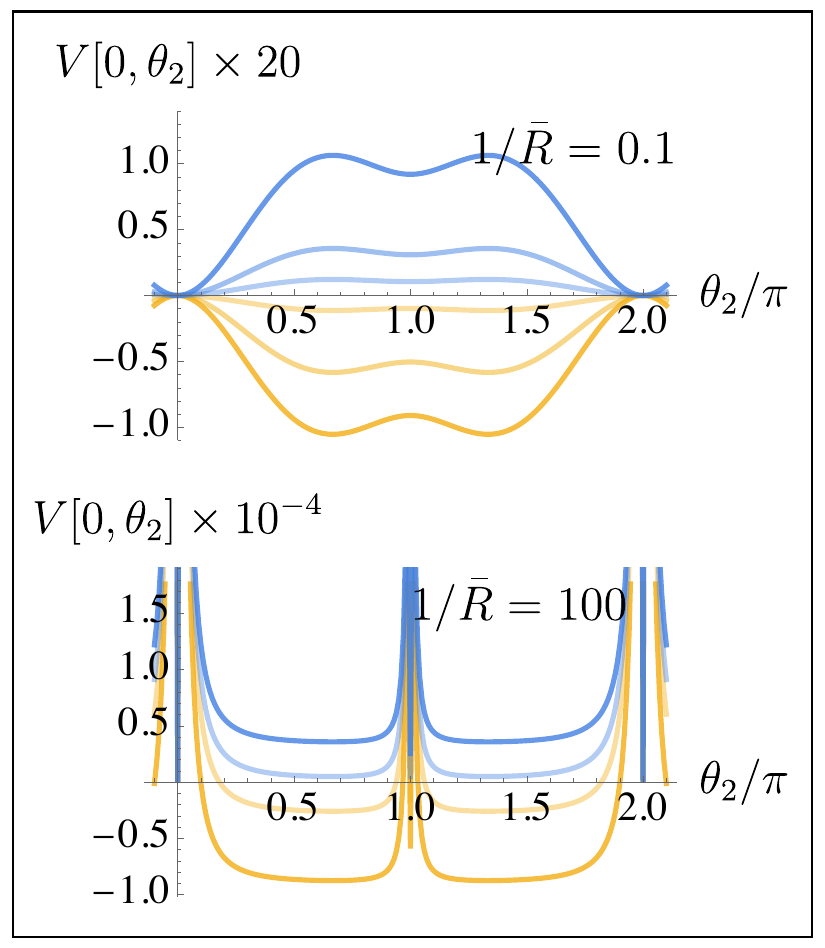} 
	
	\caption{ On the left-hand side plot, the AB phase diagram in the presence of one adjoint fermion $(N_{\rm f},N_{\rm ad})=(0,1)$ as a function of the compactification radius and the adjoint fermion mass. Three phases are found to appear: deconfined (($A$), in blue), split (($B$), in green) and reconfined (($C$), in orange). On the first set of plots on the right-hand side panel, we show the evolution of a slice of the effective potential along a fixed $\bar m_{\rm ad}=0.4$  for different finite compactification radii. In the top plot, we show the evolution from the deconfined (blue lines) to the split phase (green lines) and in the bottom plot, from the split (green lines) to the reconfined phase (yellow lines). We slightly vary the compactification radius in the interval $1/ \bar R =[0.42,0.66]$.
	In the right-most set of plots, we show the evolution of the gauge potential at two fixed radii $1/\bar R =\{0.1,10^{2}\}$. The adjoint fermion mass is slightly decreased showing the evolution of the potential from the deconfined (dark blue lines) to the split phase (dark yellow lines) in both dimensional limits. While in the $D=5$ limit (top plot), this phase transition displays a separating barrier, in the $D=4$ limit (bottom plot) the potential undergoes a flat shape.  }
	\label{Fig:potential phases 1 adjoint fermion}
\end{figure*}
%%%%%%%%%%%%%%%
%%%%%%%%%%%%%%%

As a first approximation, we neglect the effect of the anomalous dimensions in Eq.~\eqref{eq: Flow of the potential} leading to an effective potential independent of the gauge coupling. Moreover, this implies that the confined phase (denoted by $X$) cannot be accessed in the current simplified approximation. Additionally, we assume a cut-off independent fermion mass and hence its quantum corrections do not feed back to the potential. This approximation neglects higher order loop effects which become sizeable in the strong coupling limit where rich dynamics are expected.

In Fig.~\ref{Fig:potential phases 1 adjoint fermion}, we investigate the gauge vacuum phase diagram as a function of the compactification radius and the mass of one adjoint fermion.  As previously known from Monte-Carlo simulations based on Lattice gauge field theory~\cite{Cossu:2009sq,Cossu:2013ora} and effective-model studies~\cite{Kashiwa:2013rmg,Nishimura:2009me}, three phases are present in both dimensional limits. While for large masses the fermion contributions decouple and hence the deconfined phase is present, for small masses the reconfined phase is always found. 
On the right-hand side of Fig.~\ref{Fig:potential phases 1 adjoint fermion}, the left-most set of plots shows the relevant slice of the effective potential  $(\theta_1=0)$ at  a fixed adjoint fermion mass of $\bar m_{\rm ad}=0.4$ and for different compactification radius. On the top plot, the compactification radius is slightly decreased from the deconfined phase (blue line) to the split phase (green lines). In the bottom plot, we show the evolution from the split to the reconfined phase (yellow lines).

In comparison to previous studies~\cite{Kashiwa:2013rmg,Nishimura:2009me}, in the present work we can continuously connect from deeply in both $D=4$ and $D=5$ limits. The fRG accounts for the threshold effects associated to inherent scales or field masses. As shown in Eq.~\eqref{eq: threshold function with m and theta}, the propagators contain threshold functions which are sensitive to the compactification radius and the explicit fermionic mass. We find that the split phase only appears in the inter-dimensional limit (finite $\bar R$) while in the $D=4,5$ limits vanishes as is squeezed by both deconfined and reconfined phases. In the right-most panel of Fig.~\ref{Fig:potential phases 1 adjoint fermion}, we show the gauge potential evolution as the adjoint fermion mass is decreased in both of the dimensional limits. We see how the split phase (minimum at $(\theta_1,\theta_2)=(0,\pi)$) is not realised in none of limits. While in the $D=5$ limit the potential flattens along this transition, in the $D=4$ limit a barrier separates all three phases is present. Additionally, we note the clear relation between the profile of the phase boundaries with the dependence of the effective dimensionality on the compactification radius.

We also investigated the AB phase diagram for larger numbers of adjoint fermions. While the dependence on the compactification radius was found to be exactly the same, larger adjoint fermion masses are necessary in both dimensional limits for the transition between the deconfined and reconfined phases to occur. In other words, the profile shown in Fig.~\ref{Fig:potential phases 1 adjoint fermion} shifts to larger masses.

%%%%%%%%%%%%%%%
%%%%%%%%%%%%%%%
\begin{figure*}[ht!]
	\begin{minipage}{0.45\textwidth}
	\includegraphics[width=\columnwidth]{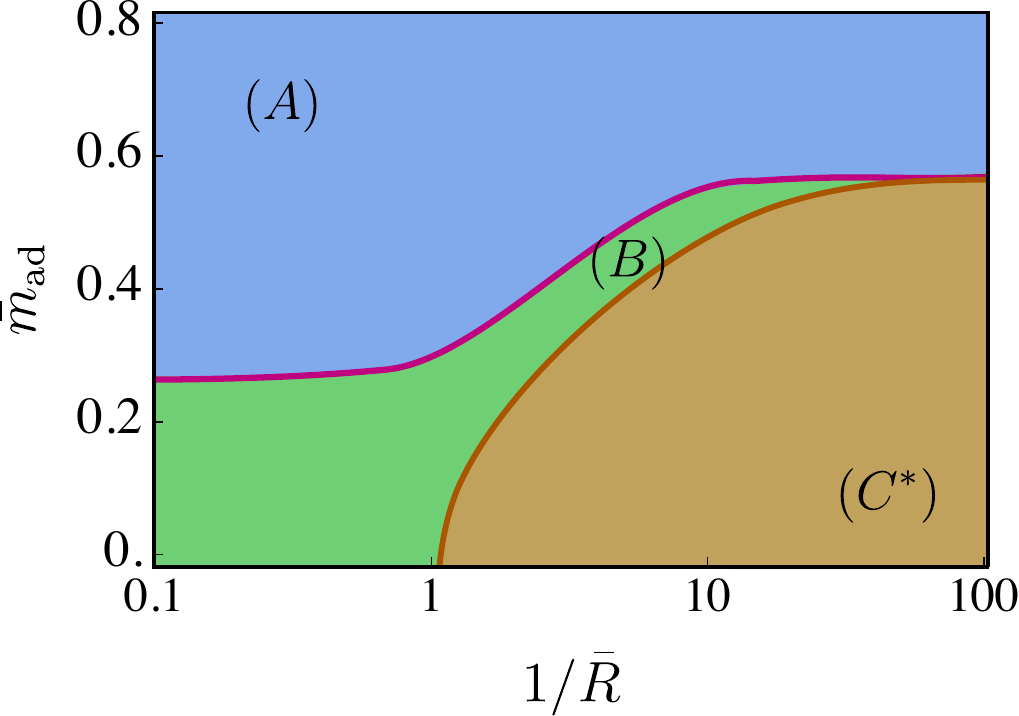}
	\end{minipage}
	\hspace{.5cm}
	\begin{minipage}{0.45\textwidth}
	\includegraphics[width=\columnwidth]{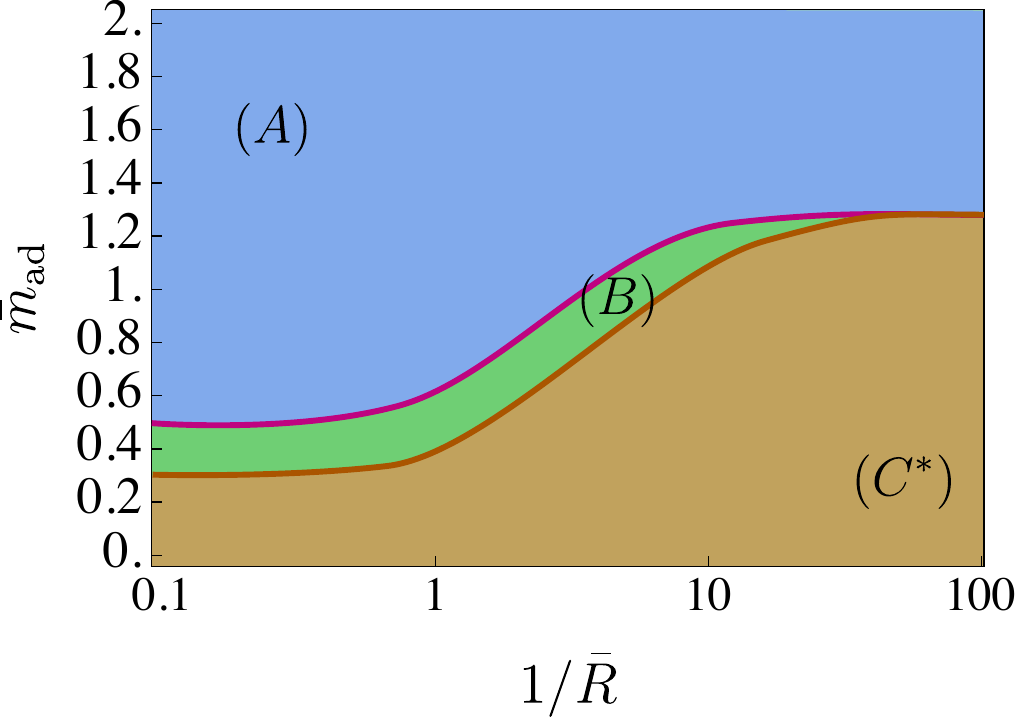}
	\end{minipage}
	\caption{ AB phase diagram as a function of the compactification radius and $\bar m_{\rm ad}$. On the left (right) hand side plot, we show the phase diagram for $(N_{\rm f},N_{\rm ad})=(1,1)$ ($(N_{\rm f},N_{\rm ad})=(2,2)$) with periodic boundary conditions and $m_{\rm f}=0$. In both cases, three phases are found: deconfined (($A$), in blue), split (($B$), in green) and pseudo-reconfined (($C^*$), in brown).
}
	\label{Fig:fundamental and adjoint}
\end{figure*}
%%%%%%%%%%%%%%%
%%%%%%%%%%%%%%%

On the other hand, the inclusion of fundamental fermions is known to explicitly break the centre symmetry of the gauge group; see e.g.~Refs.~\cite{Fukushima:2010bq,vanEgmond:2023lfu}. As a result, the gauge potential is always found in the deconfined phase ($P\neq 0$). However, including additional adjoint matter leads to a very rich phase structure. In the left plot of Fig.~\ref{Fig:fundamental and adjoint}, we show the gauge potential phase diagram for $(N_{\rm f},N_{\rm ad})=(1,1)$ with periodic boundary conditions ($\beta_{\rm f}=\beta_{\rm ad}=0$) and $\bar m_{\rm f}=0$, as a function of the compactification radius and for different values of the adjoint fermion mass. The plots here shown are the  interpolated version of the numerical results obtained. In the $D=5$ limit ($1/R\lesssim1$), we identify predominantly two phases: the deconfined and split. In between both ($\bar m _{\rm ad}\sim0.25$), we find a region in which the potential shows a flat shape in the $(\theta_1=\theta_2,\theta_2)$ plane in which both minima coexist, alike in the top plot in the right-most panel of Fig.~\ref{Fig:fundamental and adjoint}. This evolution along decreasing $m_{\rm ad}$ resembles a second order phase transition.  For small masses in the $D=4$ limit, we find the so-called ``pseudo-reconfined" phase $(C^*)$  \cite{Kashiwa:2013rmg}. In this phase (not present in Tab.~\ref{Table: summary of phases}), the Polyakov loop is slightly shifted from the central values ${\rm Im}\,P=0$ and ${\rm Re}\,P=0$ but the minimum of the potential still satisfies those conditions of the reconfined phase. 

We have also investigated the phase diagram for $(N_{\rm f},N_{\rm ad})=(2,2)$, this is shown in Fig.~\ref{Fig:fundamental and adjoint}. While the  $D=4$ limit is qualitatively similar to the $(N_{\rm f},N_{\rm ad})=(1,1)$, the  $D=5$ limit shows the presence of the pseudo-reconfined phase at small masses. Moreover, increasing the fermion number lifts the split phase to larger values of $\bar m_{\rm ad}$  and triggers the appearance of the pseudo-reconfined phase at smaller ones. This observation was already pointed out in Ref.~\cite{Kashiwa:2013rmg}. In other words, increasing fermion numbers are equivalent to considering parity-pair~\cite{Maru:2006ej} e.g. on $Z_2$ orbifolding ($S^2/Z_2$). 

%All the results discussed in the present section are in qualitative agreement with the ones found in effective field approaches~\cite{Kashiwa:2013rmg,Nishimura:2009me}. 

\section{Summary and Conclusions}
\label{sec: Summary}

In this work we have studied the properties of five-dimensional Yang-Mills theories coupled to Dirac fermions. To tackle the non-perturbative character of these theories, we have employed the functional renormalisation group to derive the flow equations for the gauge coupling and the effective potential. 

First, we have investigated the structure of the flow equations and the fixed-point solutions of the four- and five-dimensional setups with different fermionic content. Considering the extra-dimension to be compactified and employing a mass-dependent renormalisation scheme allows to continuously connect the phase diagram in both dimensional limits. In Fig.~\ref{Fig: UV fixed points regimes in change of R},  we summarise the IR and UV fixed-point solutions found at different compactification radii. Particularly, we find theories in which both IR and UV fixed points are simultaneously present. We display the trajectories between fixed points considering different onsets of the extra-dimension.

We also studied the gauge potential and AB phase diagram as a function of the compactification radius. We particularised on $N_c=3$ theories with both, adjoint and fundamental fermions. The findings discussed are in agreement with previous results from Monte-Carlo simulations based on Lattice gauge field theory~\cite{Cossu:2009sq,Cossu:2013ora} and effective-model studies~\cite{Kashiwa:2013rmg,Nishimura:2009me}. However, we find qualitative advances given we employ a mass-dependent renormalisation scheme with sensitivity to threshold effects. 

Compactifying the extra-dimension and defining an effective dimensionality has allowed us to continuously interpolate between the four- and five-dimensional limits of the theory. This is particularly useful for the implementation of the present theoretical results along functional constructions of the SM~\cite{Pastor-Gutierrez:2022nki,Goertz:2023pvn} to describe and test phenomenological models of Gauge-Higgs unification, see e.g.~Ref.~\cite{Hosotani:2023poh}.  

Given the intricacy of Yang-Mills theories, in the present work we have focused on the qualitatively features. In addition to the dimensionality of the gauge coupling, five-dimensional gauge theories also require a non-perturbative treatment as the gauge interactions turn strong. In this limit important phenomena such as dynamical chiral symmetry breaking or colour confinement take place. Here, we have worked in the background field approximation and used heat-kernel techniques to derive the flows. This approximation does not provide quantitative access to the strong sector and hence we focused in the discussion of the qualitative aspects. However,  to improve the current approximation, we have made use of three-loop $\overline{\rm MS}$ results accounting for high order fermionic contributions. This allowed to track the non-trivial CBZ fixed-point solution in five-dimensional limit.

Several improvements can be made for conclusive statements and quantitative precision. First, the flow equation for the gauge coupling can be derived from a vertex expansion of the effective action and from the scaling of 3- and 4-point functions containing gauge fields. This method facilitates the study considering higher dimensional operators. This step is crucial in order to determine the dimensionality of the critical surface and hence the predictivity of the theory.  Both analysis, the fixed point landscape and the phase diagram, can be improved by feeding back the anomalous dimensions of the gauge, ghost and fermionic fields. In the fixed point study performed, this will lead to the inclusion of more contributions at two and higher loops which at the current point have been included by employing perturbative results. In the AB phase analysis, feeding the anomalous dimension will make the gauge potential dependent on the gauge coupling. This will lead to qualitative access to the strong coupling confinement phase.

Finally, let us briefly mention some prospects of the study of extra-dimensional Yang-Mills theories. From the theoretical point of view, the non-perturbative renormalisability, i.e. the asymptotic safety scenario for extra-dimensional Yang-Mills theories, is still an open question. Within the truncation employed in this work, the $\tr[F_{\mu\nu}F^{\mu\nu}]$ operator is found to be relevant. Nevertheless, one cannot exclude the possibility of higher dimensional operators additionally being relevant and hence theory being less predictive. To address this issue, we need to determine the dimensionality of the UV critical surface in theory space spanned by an infinite number of effective operators such as $(\tr[F_{\mu\nu}F^{\mu\nu}])^n$ and $\tr[F_{\mu\nu}F^{\nu\sigma}F_{\sigma}{}^\mu]$. This is under investigation as an ongoing project.

On the phenomenological side, UV complete extra-dimensional Yang-Mills theories are necessary to construct predictive models for Gauge-Higgs unification.  The findings here discussed are relevant for phenomenological implementations of the Hosotani mechanism into viable models. However, defining chiral fermions is crucial for realistic setups of Gauge-Higgs unification models compatible with the SM. Although one cannot define chiral fermions on $\mathbb R^4\times S^1$ spacetimes, implementations are possible on the orbifold~\cite{Pomarol:1998sd}, e.g. $\mathbb R^4\times S^1/Z_2$. Additionally, realistic Gauge-Higgs unification models are usually defined as a $SU(3)_c\times SO(5)\times U(1)_X$ gauge theory~\cite{Agashe:2004rs,Medina:2007hz,Hosotani:2008tx,Funatsu:2019fry,Funatsu:2020znj}. 
%Fermion wavefunctions in extradimensional models are studied in Refs.~\cite{Nussinov:2001rb,Girmohanta:2020llh}. 
Furthermore, Gauge-Higgs grand unification theories have been studied as BSM models~\cite{Haba:2002vc,Lim:2007jv,Kojima:2011ad,Hosotani:2015hoa,Yamatsu:2015oit,Furui:2016owe,Angelescu:2021nbp,Angelescu:2022obm,Kojima:2023mew}. For these reasons, it may be worthy to study the non-perturbative features of extra-dimensional gauge theories with other gauge groups and spacetime structures.

\quad
\subsection*{Acknowledgements}
The work of M.\,Y.\ is supported by the National Science Foundation of China (NSFC) under Grant No.~12205116 and the Seeds Funding of Jilin University.

\bibliography{refs}
\end{document}